\documentclass{aa}
\usepackage{natbib}
\usepackage{threeparttable}
\bibpunct{(}{)}{;}{a}{}{,}
\usepackage{graphicx}
\usepackage{txfonts}
\usepackage{longtable,lscape}
\begin{document}
   \title{New young planetary nebulae in IPHAS}

%   \subtitle{I. Overviewing the $\kappa$-mechanism}
   \author{K. Viironen
          \inst{1}
          \and
          A.~Mampaso\inst{1}
          \and
          R.~L.~M.~Corradi\inst{1,2}
	  \and
	  M.~Rodr{\'{i}}guez\inst{3}
	  \and
	  R.~Greimel\inst{4}
	  \and
	  L.~Sabin\inst{5}
	  \and
	  S.~E.~Sale\inst{6}
	  \and
	  Y.~Unruh\inst{6}
	  \and
	  G.~Delgado-Inglada\inst{3}
	  \and
	  J.~E.~Drew\inst{7}
	  \and
	  C.~Giammanco\inst{1}
	  \and
	  P.~Groot\inst{8}
	  \and
          Q.~A. Parker\inst{9}
          \and
	  J.~Sokoloski\inst{10}
	  \and
	  A.~Zijlstra\inst{5}
	            }

   \offprints{K. Viironen}

   \institute{Instituto de Astrof{\'{i}}sica de Canarias (IAC), C/V{\'{i}}a L\' actea s/n,
              38200 La Laguna, Tenerife, Spain\\
              \email{kerttu@iac.es}\\
         \and
	 Isaac Newton Group of Telescopes, Apartado de Correos 321, E-38700 Sta. Cruz de La Palma, Spain \\
	 \and
Instituto Nacional de Astrof{\'{i}}sica, \'Optica y Electr\'onica (INAOE), Apdo Postal 51 y 216, 72000 Puebla, Pue., Mexico\\
	 \and
Institut f\"ur Physik, Karl-Franzens Universit\"at Graz, Universit\"atsplatz 5, 8010 Graz, Austria\\
 \and
	 Jodrell Bank Centre for Astrophysics, Alan Turing Building, University of Manchester, Manchester, M13 9PL, UK\\
	 	 \and
	 Imperial College London, Blackett Laboratory, Prince Consort Road, London SW7 2AZ, UK\\
	 	 \and
	 Centre for Astrophysics Research, University of Hertfordshire, College Lane, Hatfield AL10 9AB, UK\\
	 \and
	 Department of Astrophysics/IMAPP, Radboud University Nijmegen, PO Box 9010, 6500 GL, Nijmegen, the Netherlands\\
         \and
	 Department of Physics, Macquarie University, Sydney, Australia\\
	 	 \and
	 Columbia Astrophysics Lab., 550 W120th St., 1027 Pupin Hall, MC 5247, Columbia University, New York 10027, USA\\
	      }

   \date{}

% \abstract{}{}{}{}{}
% 5 {} token are mandatory

  \abstract
   {}
    {We search for very small-diameter galactic planetary nebulae (PNe) representing the earliest phases of PN evolution. The IPHAS catalogue of H$\alpha$-emitting stars provides a useful basis for this study since all sources present in this catalogue must be of small angular diameter.}
    {The PN candidates are selected based on their location in two colour-colour diagrams: IPHAS ($r^\prime  - H\alpha$) vs. ($r^\prime  - i^\prime$), and 2MASS ($J - H$) vs. ($H - K_s$). Spectroscopic follow-up was carried out on a sample of candidates to confirm their nature.}
    {We present a total of 83 PN candidates. We were able to obtain spectra or find the classification from the literature for 35 candidates.  Five of these objects are likely to be new PNe, including one large bipolar PN discovered serendipitously close to an emission-line star. PN distances deduced from extinction-distance relations based on IPHAS field-star photometry are presented for the first time.  These yield distance estimates for our objects in the range 2~kpc and 6~kpc. From the data in hand, we conclude that four of the discovered objects are probably young PNe.}
    {}


   \maketitle
%
%________________________________________________________________

\section{Introduction}\label{s1}

We present a search for planetary nebulae (PNe) of very small angular size. These objects are especially interesting because they are likely to be very young, and can therefore give us information about the late stage in the evolution of low and intermediate-mass stars. The formation of PNe and their wide variety of morphologies, arising from initially spherically symmetric AGB stars, is particularly problematic. In the past few years, high-resolution imaging of late AGB stars, proto-PNe (i.e., objects in which the central star large-scale mass-loss has turned off, but the star has not become hot enough to ionise the surrounding ejecta), and young PNe has shown that the asymmetries are present already {\it before} the planetary nebula phase \citep{sahai07}. This is not explained by the classical PN formation model \citep[i.e., the Generalised Interacting Stellar Winds model,][]{kwok82}.

To solve this problem, it is essential to carry out detailed studies of proto-PNe, transition objects, and young PNe (e.g. \cite{suarez06} and \cite{sanchez-contreras08}). The youngest PNe are expected to be very dense and of small size and there is a paucity of very small nebulae in the catalogues. For instance, a major PNe catalogue \citep[Strasbourg,][]{acker94} lists only one object with a measured optical diameter of less than 1 arcsec, which, in fact, is not a young PN but the halo PN G061.9+41.3. At radio wavelengths, there are some 25 PNe with subarcsec diameters but they mostly correspond to compact radio cores in optically-extended nebulae. All other entries with an optical size noted as stellar in the Strasbourg catalogue represent limits where the size could not be measured, clearly reflecting the limited resolution of the old (often photographic) imaging surveys for PNe. Even the MASH catalogue of PNe \citep{parker06,miszalski08} is limited to $\sim3\arcsec$ for the smaller PNe. 

\citet{sahai07} studied 23 proto-PNe with the Hubble Space Telescope, measuring apparent sizes from 0.1 to 25$\arcsec$ (typically 1$\arcsec$): their spectra, when available, lack many of the strong emission lines typical of mature PNe, consistent with a proto-PN (rather than  young-PN) designation. \citet{cerrigone08} mapped a sample of very young PNe at radio wavelengths, finding typical diameters between 0.3 and 1.2$\arcsec$, and high densities from 4$\times 10^4$ to 40$\times 10^4$ cm$^{-3}$. These are the type of PNe we are searching for in the optical.

IPHAS (The INT/WFC Photometric H$\alpha$ Survey of the Northern Galactic Plane) provides a new database to search for very compact PNe because of its depth and high spatial resolution. IPHAS is mapping all Galactic latitudes between $b$= --5 to +5 degrees, in three filters using the Wide Field Camera (WFC) on the Isaac Newton Telescope (INT) at the Observatorio del Roque de los Muchachos (La Palma, Spain). IPHAS observing started in August 2003 and is now more than 90\% complete. A narrow-band H$\alpha$ (central wavelength and width: 6568~\AA/95~\AA) and two Sloan filters ($r^\prime$ and $i^\prime$) are used in matched 120, 30, and 10 s exposures, respectively, spanning the magnitude range $r^\prime$=13 to 21 mag for point sources. Each IPHAS field is observed twice at two closely overlapped pointings. More information about the survey and its database can be found in \cite{drew05} and \cite{gonzalez-solares08}.

\cite{witham08} (hereafter W08) published a catalogue of 4853 high-confidence H$\alpha$ emission-line stars selected from IPHAS. \cite{corradi08} showed that most of the objects in the catalogue are Be stars and that different kinds of emission-line objects can be separated from each other based on their IPHAS and 2MASS colours. They further searched for symbiotic stars, discovering 1183 candidates, and presented the spectroscopic confirmation of three new symbiotic stars. We base our PN search on the W08 catalogue and expect to find only point-like or barely resolved PNe.

In Section~\ref{s2} of the paper we explain our selection methods; in Section~\ref{s3}, the search results are presented; and in Section~\ref{s4}, the observations. In Section~\ref{s5}, individual studies of the newly found planetary nebulae are reported; in Section~\ref{s6}, their location in various diagnostic diagrams is studied; in Section~\ref{s7}, we discuss the nature and characteristics of the new objects found; and in Section~\ref{s8}, their distances are estimated. In Section~\ref{s9}, final discussion and colnclusion are provided.

\begin{figure*}
   \centering
   \includegraphics[width=0.45\linewidth,angle=0]{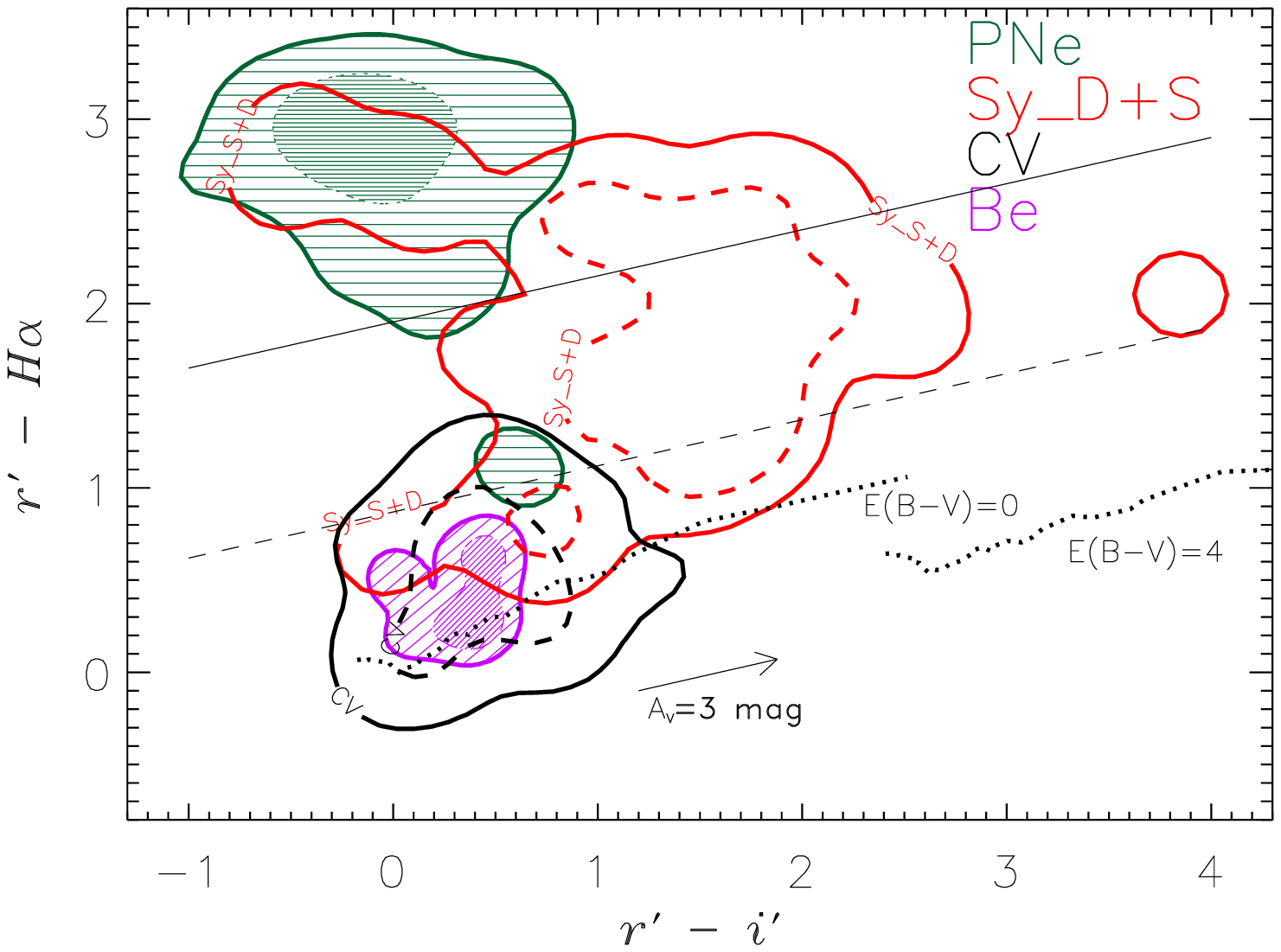}
   \includegraphics[width=0.45\linewidth,angle=0]{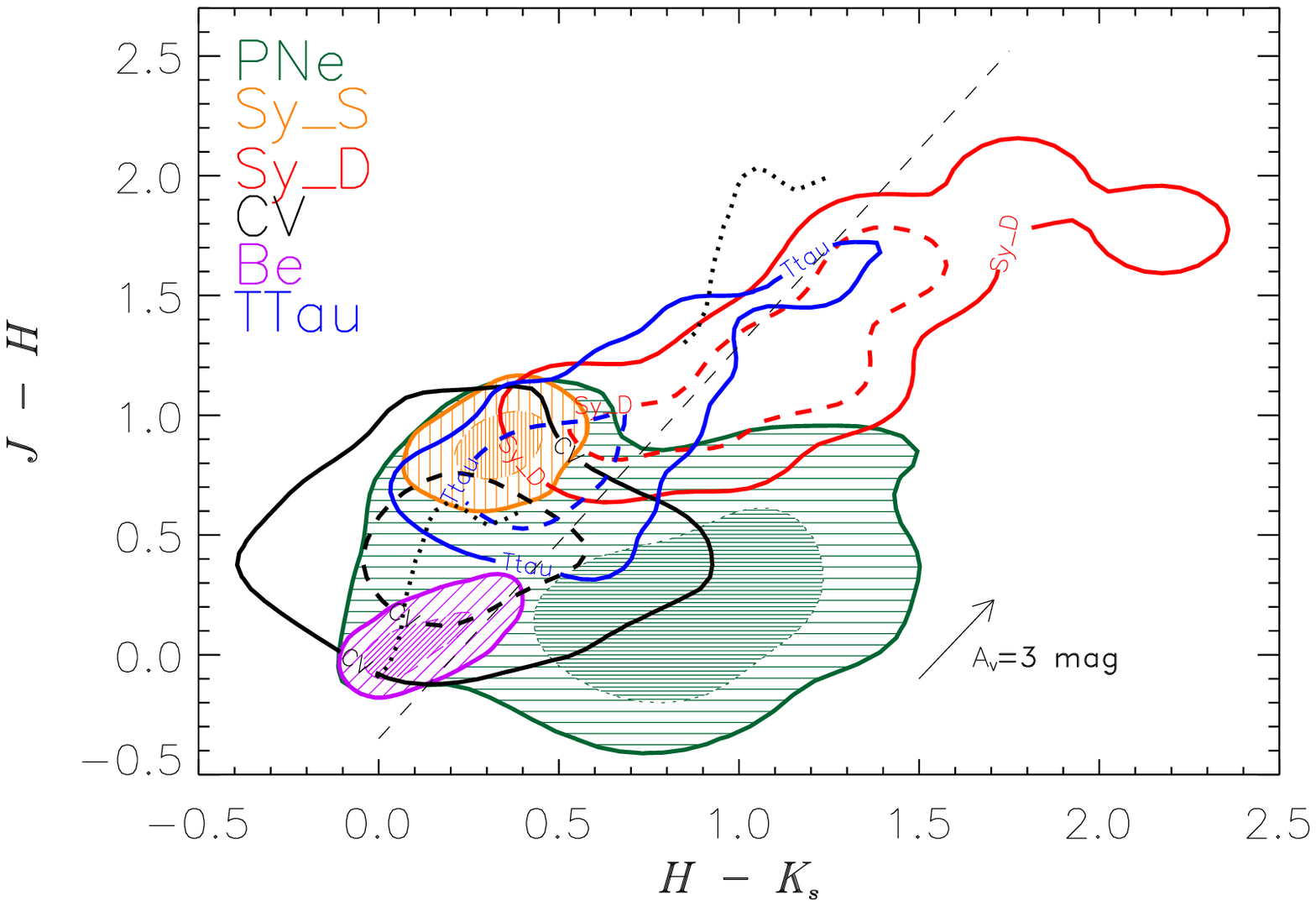}
      \caption{{\it Left} Sketch of the location of PNe (area filled with green horizontal lines), D- and S-type symbiotic stars (Sy\_D+S, red lines), cataclysmic variables (CV, black lines), and Be-stars (area filled with diagonal lines coloured magenta) in the IPHAS two-colour diagram. The location of main-sequence stars \citep{drew05} with reddenings of $E(B-V)= 0$ and $E(B-V)=4$ are shown as dotted lines. Shown are also the two cuts used in the selection of PN candidates, as explained in the text. {\it Right} 2MASS two-colour diagram for the same objects and the same MS reddening tracks as in the left. As is the available sample of symbiotic stars is larger, they are divided into S-type (Sy\_S, areas filled with vertical orange lines) and D-type (Sy\_D, red lines). The location of T-Tauri stars (TTau, blue lines) is also shown. The dashed line marks the cut used in the selection. See text for more details.}
\label{fig:1}
\end{figure*}

\begin{figure*}
   \centering
   \includegraphics[width=0.35\linewidth,angle=-90]{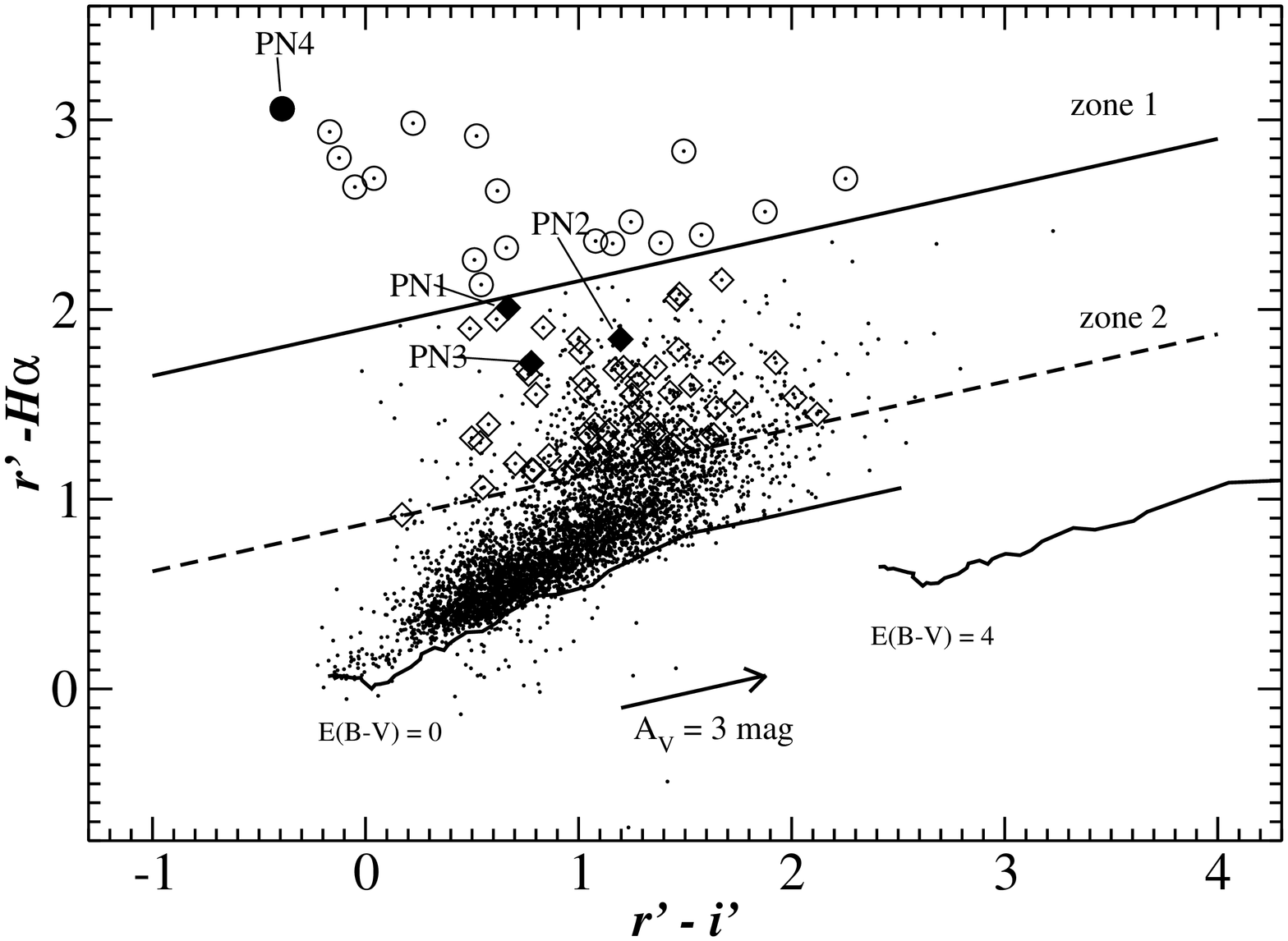}
   \includegraphics[width=0.35\linewidth,angle=-90]{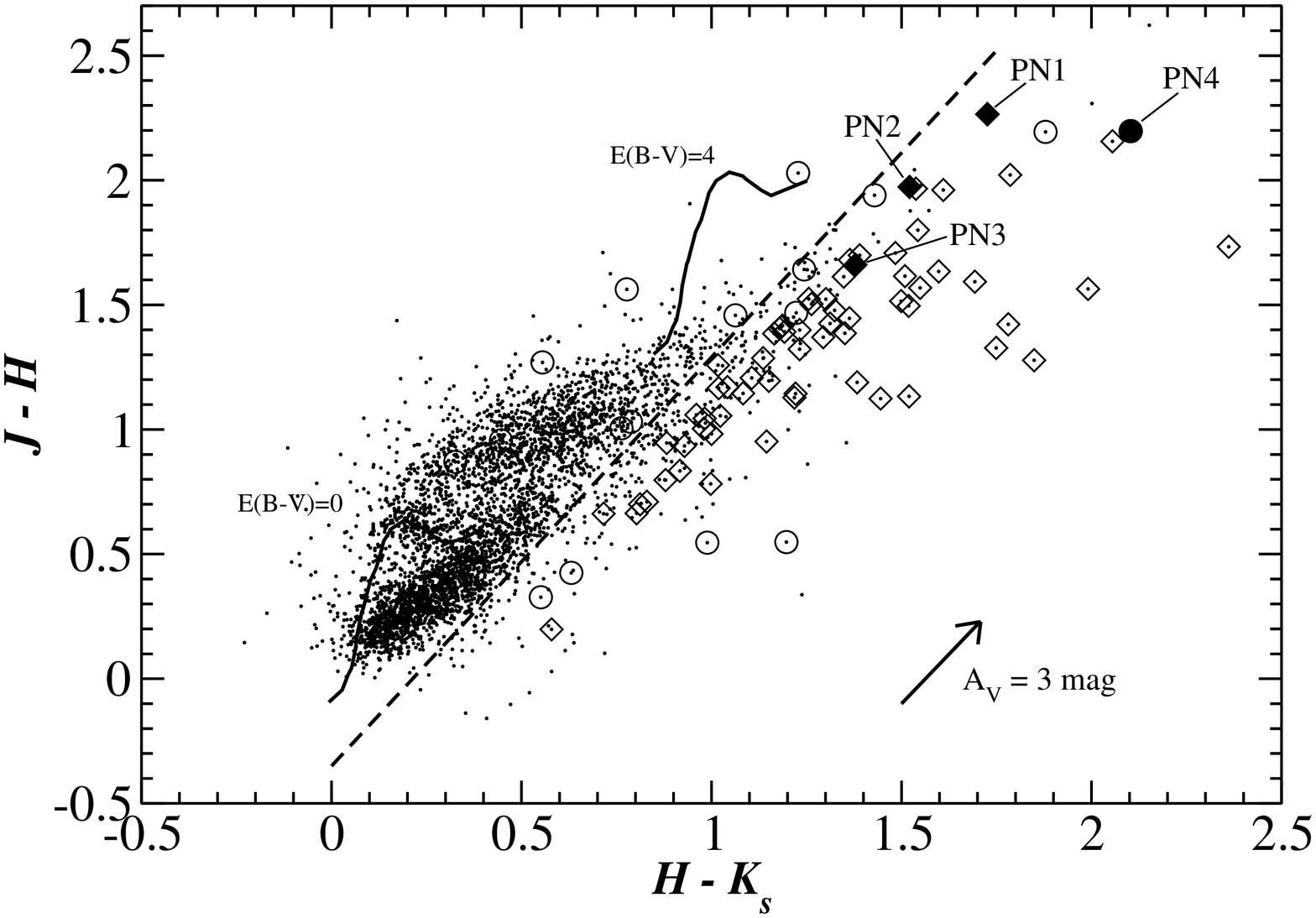}
      \caption{Left: IPHAS colour-colour diagram; {\it dots} 4853 objects from W08 catalogue, {\it circles} PN candidates from this work selected based on their IPHAS colours, and {\it diamonds} PN candidates from this work selected based on their IPHAS$+$2MASS colours. The location of main-sequence stars with reddenings of $E(B-V) = 0$ and $E(B-V) = 4$ are shown as solid lines, as marked in the figure. The two cuts used in the selection of PN candidates are also shown, as explained in the text. Right: 2MASS colour-colour diagram; the symbols used are the same as in the image at left. The dashed line marks the cut used in the selection. In both diagrams, the four new PNe discovered due to their colours are marked as well as the four known PNe included into the W08 list. Interestingly, all the three objects selected based on their IPHAS colours, and located clearly above the selection line in 2MASS are new IPHAS symbiotic stars \citep[][Corradi et al., in preparation]{corradi08}.}
\label{fig:2}
\end{figure*}

\addtocounter{table}{1}

%__________________________________________________________________

\section{Selection methods}\label{s2}

Our search for PNe is based on two colour-colour diagrams: IPHAS ($r^\prime - H\alpha$) vs. ($r^\prime - i^\prime$) and 2MASS ($J - H$) vs. ($H - K_s$). To study the position of PNe in these diagrams, the IPHAS and 2MASS photometry for all known PNe detected by these two surveys was collected and the objects were placed in both two-colour diagrams. These positions were compared with the position of other classes of objects and it was found that the diagrams can be used to separate PNe quite well from other types of emission-line and normal stars. The diagrams were discussed in \cite{corradi08} and in Fig.~\ref{fig:1} we represent a schematic view of them. The dark shaded surfaces or the dashed contours show the areas where 50\% of the objects of each kind are located, while the brighter surface and solid lines mark the areas that include 90\% of the corresponding objects.
The T-Tauri stars are not plotted in the IPHAS diagram since they spread over much of the diagram above the sequence of the normal stars \citep[see][]{corradi08}.

As can be seen, in the IPHAS diagram the PNe are well separated from the normal stellar locus and the overlap with symbiotic stars is only partial. In the 2MASS diagram, most of the known PNe are separated as well, although some lie in a confused area also occupied by other emission-line objects and some normal stars.

Figure~\ref{fig:2} illustrates our compact/barely resolved PN search in the W08 catalogue using IPHAS (left) and 2MASS (right) colour-colour diagrams. All objects from the W08 list are shown as black dots. We performed the PN candidate selection in two distinct ways:-

(1) We selected all the objects located above the straight solid line in the IPHAS diagram (19 objects). The line delimits the part of the diagram including 95\% of the known PNe and was extended parallel to an approximate reddening vector \citep[see][]{corradi08} so as to allow highly reddened objects to be selected as well. From now on, we refer to this as zone 1.

(2) For the area in the IPHAS diagram where overlap with other emission line objects is more important (below zone 1), we used 2MASS data to achieve further discrimination.  All objects located in the area including the final 5\% of the known PNe were first selected (zone 2, the area between the dashed and solid lines in IPHAS diagram in Fig.~\ref{fig:2}, left), but of these, only those located to the right of the dashed line in the 2MASS diagram were selected as PN candidates (64 objects): this cut includes 80\% of the known PNe in the 2MASS colour-colour diagram and again runs parallel to the reddening vector, permitting the selection of highly reddened candidates. For this reason, we also did not define an upper limit to $J - H$.  We do not place the cut any closer to the strip occupied by main-sequence objects, since within this narrow zone the overlap with other classes of emission-line object is very high (see Fig.~\ref{fig:2}, right).

The detection limit of the 2MASS point-source catalogue is $\sim$ 15 mag in the $K_s$ filter, while the IPHAS limit in the W08 catalogue is $r^{\prime}=19.5$ mag. Of the known PNe (those listed in either the Strasbourg or the MASH catalogues), 49 have well determined IPHAS and 2MASS magnitudes. The average $r^\prime - K_s$ colour of these PNe is 3.8. Therefore, using 2MASS as an additional selection criterion might affect the completeness of the compact PN search within IPHAS. However, considering that we can expect many of the new PNe found from IPHAS to be more reddened (and thus optically fainter) than the known ones used in calculating the observed $r - K_s$ colour, this problem becomes less important.

As can be seen in Fig.~\ref{fig:1}, we expect our selection of PN candidates to  inevitably identify other emission-line objects, such as symbiotic stars and T-Tauri stars, which are possible contaminants of our selection. For this reason, we studied the IPHAS images of every selected object: we measured the angular FWHM in H$\alpha$ and compared it with that of nearby field stars of equal brightness to study if any candidates are extended; we also studied their H$\alpha$ environment to check for possible associated diffuse emission.  Finally, we checked if the literature already contains data for any object selected.

\section{Search results}\label{s3}

The results from our search are shown in Table~\ref{tab:1}. We selected a total of 83 PN candidates, Table~\ref{tab:1} lists the following data for each: the coordinates (the external precision of IPHAS with respect to 2MASS is 0.1$\arcsec$, and the internal precision is generally higher than this); IPHAS and 2MASS photometry; IPHAS mean on-sky FWHM and its error derived from field stars of equal brightness ($\pm$0.1~mag, or a minimum of 50 stars located on the same CCD within a radius of 10\arcmin), and on-sky FWHM for the candidate, all in units of arcsecs; Simbad information; our follow-up spectroscopy (S, ''y'' meaning that a spectrum has been taken and ''n'' $=$ no spectrum); tentative object type (Id.); and the mean distance $d_4$ (in arcmin) from the candidate's four nearest neighbours listed in W08, useful for deciding if the object is likely to belong to a cluster, as is usually true for young stellar objects (YSO): smaller values of $d_4$ indicate higher probability of YSO nature for the candidate \citep[see,][]{corradi08}. For some objects, additional notes are provided.

We carefully studied the literature for the 17 objects noted in W08 to match objects classified in Simbad as PNe: most of them turned out not to be PNe. In summary, four are known PNe: PN~K~3-15, PN~K~3-18, PN~K~3-33, and PN~K~3-49, and these are plotted in Fig.~\ref{fig:2}. Three are listed as possible PNe (no confirmation found in the literature): J050327.55+414217.3, J191528.59+184748.3, and J191750.56+081508.5. Study of the literature data indicates that the remaining 10 objects are not PNe (they are, e.g., H$\alpha$ emission stars, H\,{\sc ii} regions). Our method selected the four known PNe listed above and one of the possible PNe, while the other two possible PNe and the ten non-PNe were not selected. Of the known PNe, only one (PN~K~3-18) is extended in H$\alpha$ (0.7$\arcsec$ deconvolved FWHM). In addition, PN~K~3-33 shows an unresolved nucleus and extended emission in its surroundings.

We either obtained spectra or the object type is determined from the literature for 35 of the candidates. The analysis of the spectra of the objects that do not show the typical PN lines in emission is in progress and will be published later (Valentini et al., in preparation). These  objects are simply marked as emission-line stars, Em*, in Table~\ref{tab:1}. Besides the four known PNe discussed above, we discovered another four objects that are likely to be young compact PNe, as confirmed by our follow-up spectroscopy: IPHAS J012544.65+613611.6, IPHAS J053440.77+254238.2, IPHAS J194907.22+211741.9, and IPHASX J200514.59+322125.1. For brevity, from now on we refer to them PN1, PN2, PN3, and PN4, respectively. We present these objects in the next section. In addition, we present the new extended PN, IPHASX J221117.99+552841.0 (PN5), which was discovered serendipitously because of its proximity to a W08 object.

\section{Observations of the new PNe}\label{s4}

In addition to IPHAS imaging, we carried out follow-up imaging for the extended new PNe and spectroscopy for all the new PNe.

\subsection{IPHAS imaging}

\begin{figure}
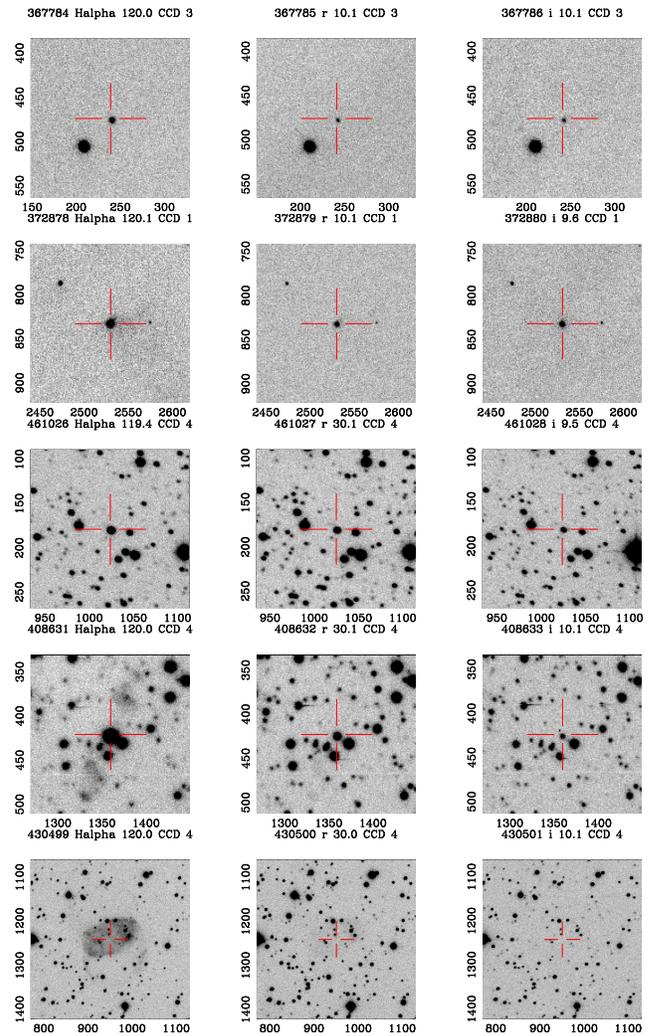

\centering
\includegraphics[width=0.3\linewidth,angle=-90]{Fig3a.ps}
\includegraphics[width=0.3\linewidth,angle=-90]{Fig3b.ps}
\includegraphics[width=0.3\linewidth,angle=-90]{Fig3c.ps}
\includegraphics[width=0.3\linewidth,angle=-90]{Fig3d.ps}
\includegraphics[width=0.3\linewidth,angle=-90]{Fig3e.ps}
\protect\caption[ ]{IPHAS H$\alpha$, r$^{\prime}$, and i$^{\prime}$ (left, centre, right column, respectively) images of PN1, PN2, PN3, PN4, and PN5, from top to bottom. In all images, north is up and east towards left. The box dimensions are $60\arcsec \times 60\arcsec$ except for PN5, for which the dimensions are $120\arcsec \times 120\arcsec$.
\label{fig:3}}
\end{figure}

Information about the IPHAS images for the new PNe is given in Table~\ref{tab:0}: we list the IPHAS name of the object, the nickname used in this paper, observing date(s), average seeing for the corresponding IPHAS field (in arcsecs), and the number of available IPHAS image sets (H$\alpha$, $r^\prime$, $i^\prime$) in which the PN appears. Fig.~\ref{fig:3} shows the IPHAS images.

\tiny{
\begin{table}
\caption{\label{tab:0} IPHAS imaging for the new PNe.}
 \centering
 \begin{tabular}{l l l l l}
 \hline \hline
Object & abbr.  &  date & seeing & n\\
\hline
\tiny{IPHAS J012544.65+613611.6} & \tiny{PN1} & \tiny{20031013} & \tiny{1.3} & \tiny{4}\\
\tiny{IPHAS J053440.77+254238.2} & \tiny{PN2} & \tiny{20031105} & \tiny{0.7} & \tiny{2}\\
\tiny{IPHAS J194907.22+211741.9} & \tiny{PN3} & \tiny{20050613} & \tiny{1.0} & \tiny{2}\\
\tiny{IPHASX J200514.60+322125.4} & \tiny{PN4} & \tiny{20040711,26} & \tiny{1.4,1.7} & \tiny{3}\\
\tiny{IPHASX J221117.99+552841.0} & \tiny{PN5} & \tiny{20041103,05} & \tiny{1.3,1.3} & \tiny{3}\\
\hline
\hline
\end{tabular}
\end{table}
}
\normalsize

\subsection{Other imaging}

\begin{figure}
   \begin{center}
   \includegraphics[width=0.6\linewidth,angle=0]{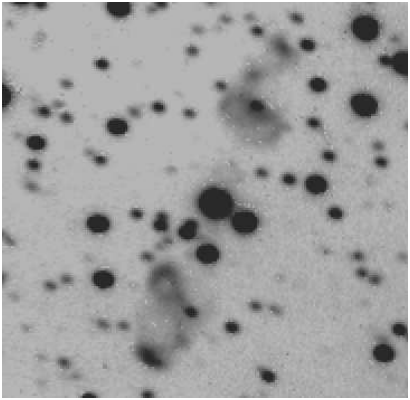}
         \caption{\label{fig:4} Image of PN4 in H$\alpha$+[N\,{\sc ii}] plotted using a logarithmic intensity scale. The size of the box is 60$\arcsec$, north is up and east is to the left.}
\end{center}
\end{figure}

In addition to IPHAS survey imaging, $3\times30$ min H$\alpha$ exposures were obtained for PN4 (Fig.~\ref{fig:4}) on 9 June 2006.  The seeing at the time was $\sim$ 1.2$\arcsec$. The instrumental set-up used was the same as used for IPHAS.

We also obtained a 93.2 ks observation of PN4 with the $Chandra$ X-ray Observatory, on 2007 January 15, using the ACIS-S/S3 detector.   Neither the central star nor the lobes produced detectable X-ray emission.  In the 0.3-10 keV energy range, the 3-$\sigma$ upper limits for the central star, the NW lobe and SE lobe are respectively  $< 1.6 \times 10^{-4}$~counts~s$^{-1}$, $< 4.6 \times 10^{-4}$~counts~s$^{-1}$, and $< 4.8 \times 10^{-4}$~counts~s$^{-1}$.

\begin{figure*}
   \begin{center}
     \includegraphics[width=0.3\linewidth,angle=0]{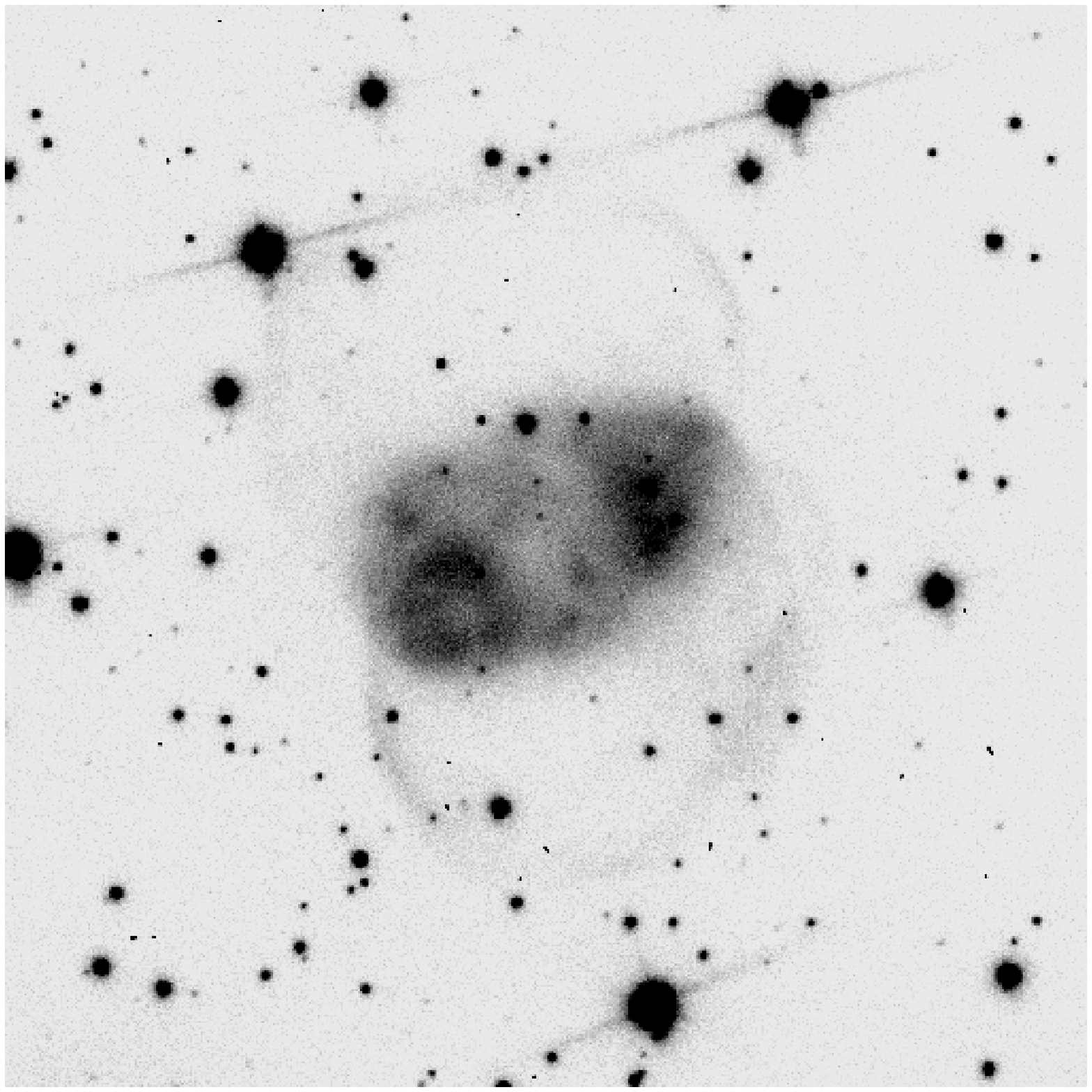}
   \includegraphics[width=0.3\linewidth,angle=0]{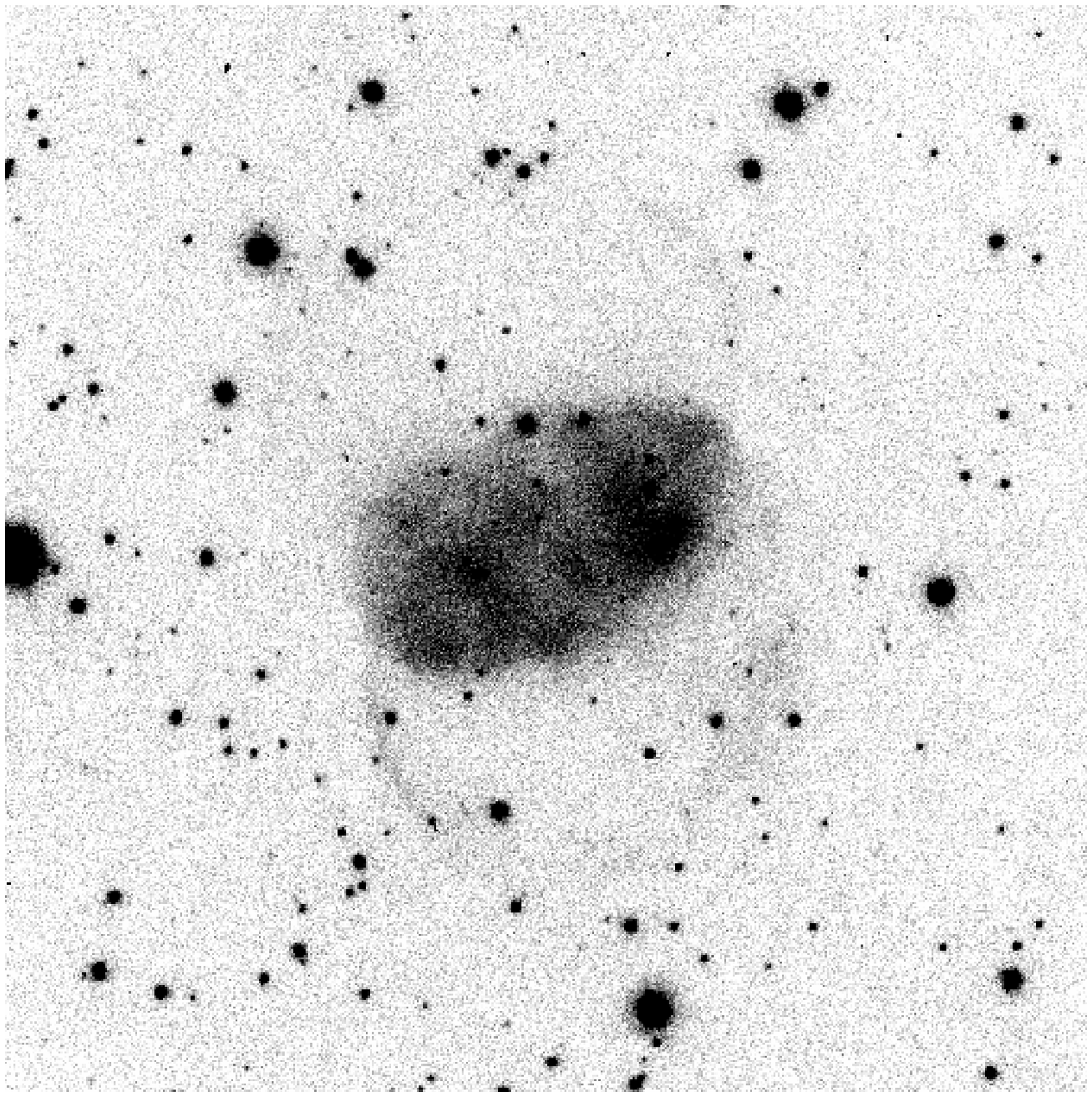}
\includegraphics[width=0.3\linewidth,angle=0]{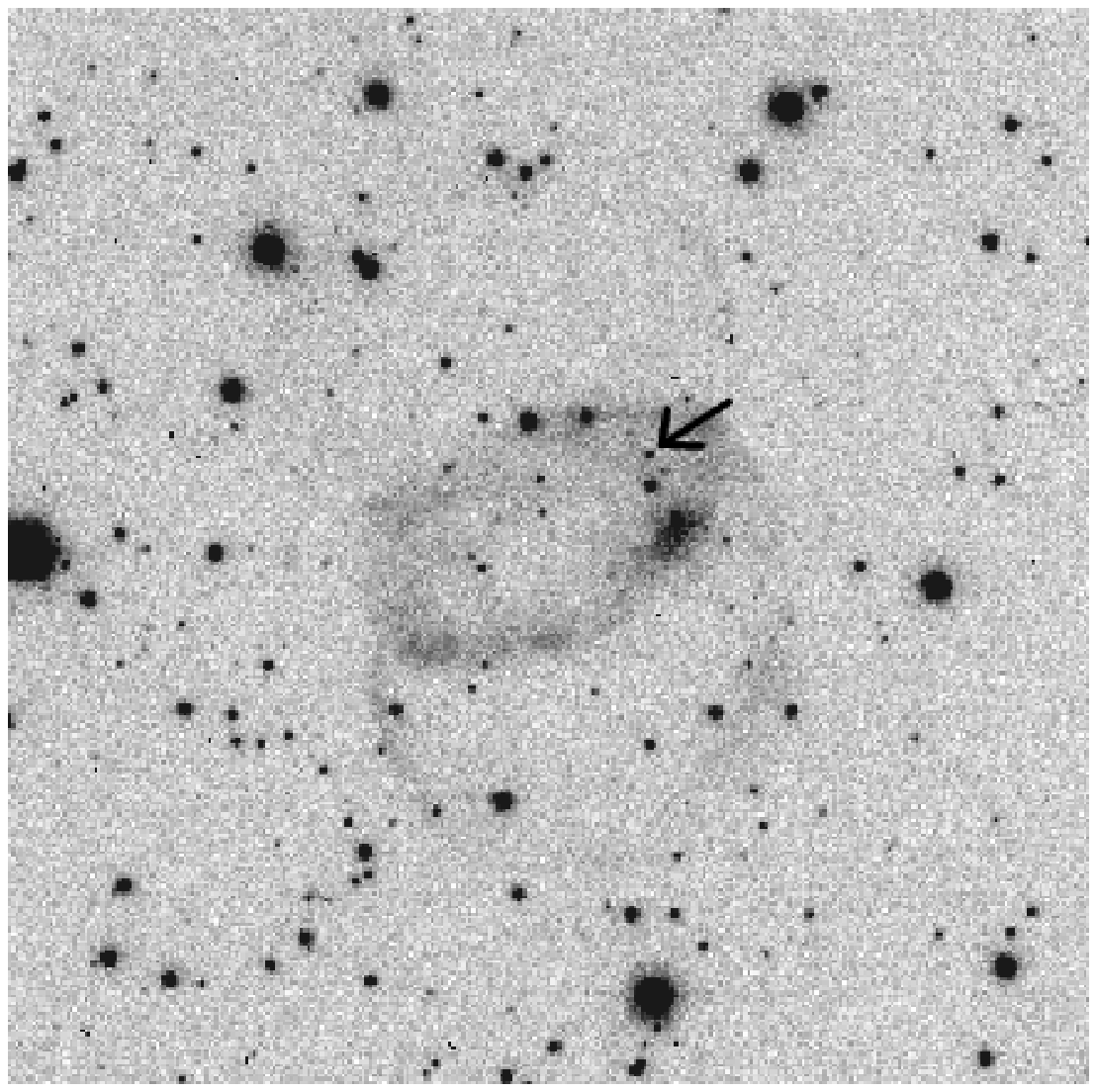}
      \caption{\label{fig:5} The NOT [O\,{\sc iii}] (left), H$\alpha$ (centre), and [N\,{\sc ii}] (right) images of PN5. The sides of the square cut-outs are 90$\arcsec$ long. North is up and east is to the left. The star selected in W08 catalogue is shown by an arrow in the [N\,{\sc ii}] image. The J2000.0 coordinates of the star are 22:11:16.75 +55:28:51.5.}
\end{center}
\end{figure*}

PN5 was observed with the Nordic Optical Telescope (ORM, La Palma) and ALFOSC (E2V 2048 $\times$ 2052 CCD) on 5 September 2007. The plate scale was 0.19 $\arcsec$pix$^{-1}$ and the seeing, $\sim$0.5$\arcsec$. 30 min [O\,{\sc iii}], H$\alpha$, and [N\,{\sc ii}] exposures were obtained  (Fig.~\ref{fig:5}). The central wavelengths and widths of the filters are 5007/3~\AA, 6564/33~\AA, and 6583/3.6~\AA, respectively.

\subsection{Spectroscopy}

The details of the spectroscopic follow-up of the new PNe are given in Table \ref{tab:sfollow}. In the observations from San Pedro M\'artir (SPM), Mexico, of PN4, the slit was positioned at P.A.= 157$^{\circ}$, thus passing through the core and the two lobes. The effective airmass at the time of observing was 1.8. When observing PN5 with the INT, the slit was positioned at P.A.= 115$^{\circ}$, thus passing through the two blobs visible in the [O\,{\sc iii}] image (see Fig.~\ref{fig:5}). The effective airmass at the time of observing was 1.5. Otherwise, the slit was always positioned to match the parallactic angle.  Bias frames, tungsten flat-field exposures, arcs, and spectroscopic standards were observed on each night of data collection.  The spectra were reduced and flux-calibrated using the standard IRAF/FIGARO packages, while line fluxes were measured by fitting Gaussian profiles using the task SPLOT of IRAF.

\begin{table*}
\begin{minipage}[t]{\linewidth}
\caption{\label{tab:sfollow} Spectroscopic observations for the new PNe presented in this paper.}
 \centering
\renewcommand{\footnoterule}{}
 \begin{tabular}{l l l l l l l}
 \hline \hline
Object & date & telescope\footnote{NOT, INT and WHT are located at the Observatorio de Roque de los Muchachos on La Palma, Spain. SPM 2.1m is located at the Observatorio Astron\'omico Nacional de San Pedro M\'artir, Mexico.} & instrument & $\lambda$ range [\AA] & dispersion [\AA~pix$^{-1}$] & exp. time [min]\\
\hline
PN1 & 20070920 & NOT & ALFOSC & 3000-9100 & 3 & 30\\
PN2 & 20060809 & INT & IDS & 3090-8750& 1.8 & 15\\
PN3 & 20070728 & WHT & ISIS (red,blue) & 3420-5750,6384-9244 & 1.7,1.9 & 6,6\\
PN4 &  20041023,20060909 &  SPM2.1m/INT & B\&Ch (red,blue)/IDS &  3720-7470/3090-8750 & 2.0/1.8 &  40,20/30 \\
PN5 & 20060828 & INT & IDS & 3090-8750 & 1.8 & 30\\
 \hline
\hline
\end{tabular}
\end{minipage}
\end{table*}

\section{Results for the new PNe}\label{s5}

In this section, the results from the imaging and spectroscopic study of the new PNe are presented.

\subsection{Morphology and environment}

All sources in IPHAS are located close to the Galactic Plane where many star-forming complexes, H\,{\sc ii} regions, and molecular clouds are present, and there is a high probability of a candidate being nearby in projection, or physically associated, with one of those regions. Therefore, it is important to study the morphology (when the object is resolved) and environment around the candidates, both to discriminate associated nebulosities from unrelated ones (like e.g., H\,{\sc ii} regions or diffuse emission regions) and to provide additional information when studying the nature of the objects.

PN1 is unresolved. Thus, assuming for its FWHM an upper limit of $\textrm{FWHM}+3\times \textrm{FWHM}_{\textrm{err}}$, where $\textrm{FWHM}_{\textrm{err}}$ is the standard deviation of the FWHMs of the field stars of equal brightness (Table~\ref{tab:1}), we can calculate a three sigma upper limit of its deconvolved diameter: $\sqrt{(\textrm{FWHM}+3\times \textrm{FWHM}_{\textrm{err}})^2-\textrm{FWHM}_{\textrm{stars}}^2}$. This gives a diameter $<$ 0.6$\arcsec$.
No trace of nebulosity is found from the IPHAS images.
PN1 was detected in the infrared by MSX (G126.9996-00.9985) at 8.28 $\mu m$ only, with a flux of 0.17 Jy, and by IRAS (IRAS01224+6120) at 25 $\mu m$ (0.65 Jy) and 60 $\mu m$ (0.87 Jy).
The object is located 41$\arcmin$ SE from the center of the very large molecular cloud [YDM97] CO 164 \citep{yonekura97}, with dimensions of $96\arcmin \times 72\arcmin$. The much smaller dark cloud LDN1324 \citep{lynds62} with a diameter of $\sim 12\arcmin$ is located 25$\arcmin$ to the East of PN1.

PN2 is unresolved ( diameter $< 0.4\arcsec$). It shows a very faint irregular nebulosity extending $\approx$ 10 arcsec south-west from the source, and a slightly brighter protrusion up to 2-3 arcsec to the NW.
Like PN1, PN2 is a faint MSX source (G181.4469-03.7710) only detected at 8.28 $\mu m$ with a flux of 0.22 Jy.
PN2 is located 10$\arcmin$ SE from the center of a small molecular cloud of radius $5\arcmin$ \citep[number 64 of][]{kawamura98}.

PN3 is unresolved ( diameter $< 0.6\arcsec$); no nearby nebulosity is detected, and no IRAS nor MSX sources are associated (but note that a brighter star located 28 arcsec SW of PN3 -partially seen in  Fig.~\ref{fig:3}- is an IRAS and MSX source). No molecular cloud or star-forming regions are catalogued nearby.

PN4 has a typical morphology of a bipolar PN (Fig.~\ref{fig:4}) in that it has an unresolved stellar core and two detached elongated lobes. The apparent long axis of PN4 is 52$\arcsec$ and the aspect ratio of the lobes (length over width) is 4.6. The morphology of this PN closely resembles those of He~2-25, 19W32, and Th~2-B, objects classified as PNe but for which \cite{corradi95} argued a possible symbiotic nature.
PN4 is projected on a very extended diffuse-emission complex where many H\,{\sc ii} regions as well as dark nebulae are listed in Simbad, but the object itself is not associated with any dark nebula. It is an IRAS source, IRAS20032+3212, with fluxes of 1.90, and 3.13 Jy at 12 and 25 $\mu m$, respectively, and a MSX source with fluxes F(8.28$\mu m$)= 1.38 Jy, F(12.13$\mu m$)= 2.07 Jy, F(14.65$\mu m$)= 1.98 Jy, and F(21.3$\mu m$)$<$ 1.8 Jy.

PN5 is composed of a bright, filled, central ellipse (a ring in the [N\,{\sc ii}] image) with two bright knots at both ends of the ellipse that are especially prominent in [O\,{\sc iii}] (Fig.~\ref{fig:5}). Two faint bipolar outer lobes are visible in all three filters. The largest extent of PN5 is 79$\arcsec$ and the outer lobes are roughly round. PN5 is not detected in the infrared (IRAS and MSX) but it is the only one of our sources detected by the NVSS radio survey with $F$(1.4~GHz)$=3.0 \pm 0.5$~mJy and dimensions of $93 \times 55 \arcsec$.

\begin{table}
\caption{\label{tab:6} Line fluxes (normalised to F(H$\beta$) = 100) for the NOT+ALFOSC spectrum of PN1. The wavelengths are in units of \AA~and the statistical errors in the flux measurements and the propagated error in the dereddened flux are given within brackets as a percentage.}
 \centering
 \begin{tabular}{@{}lllll@{}}
 \hline \hline
Line, $\lambda$ & $\lambda$~Obs. & F (Obs.) & F (Dr.)\\
 \hline
{}[O\,{\sc ii}] 3726.03+3728.8 & 3726.9 & 28.3 (24) & 96.2 (24) \\
He\,{\sc i} 3935.94 & 3934.4 & 13.2 (20) & 37.4 (20) \\
H7 3970.07 & 3968.6 & 20.3 (10) & 55.7 (10) \\
{}[S\,{\sc ii}] 4068.60 + 4076.35 & 4069.6 & 98.1 (6) & 242.3 (7) \\
H$\delta$ 4101.74 & 4100.8 & 15.3 (13) & 36.5 (13) \\
{}[Fe\,{\sc ii}] 4243.97 & 4242.7 & 12.4 (17) & 25.2 (17) \\
{}[Fe\,{\sc ii}] 4287.39 & 4286.3 & 9.5 (21) & 18.3 (21) \\
H$\gamma$ 4340.47 & 4340.0 & 32.3 (7) & 58.5 (8) \\
{}[O\,{\sc iii}] 4363.21 & 4361.4 & 11.3 (18) & 19.9 (18) \\
\tiny{(+ [Fe\,{\sc ii}] 4359.34?)} & & & \\
{}[Fe\,{\sc ii}] 4413.78 & 4413.8 & 9.5 (17) & 15.8 (17) \\
Mg\,{\sc i}] 4571.20 & 4570.9 & 8.1 (23) & 11.1 (23) \\
{}[Fe\,{\sc ii}] 4814.55 & 4812.3 & 5.5 (23) & 5.7 (23) \\
H$\beta$ 4861.33 & 4861.2 & 100.0 (2) & 100.0 (2) \\
{}[O\,{\sc iii}] 4958.91 & 4957.4 & 7.9 (28) & 7.2 (28) \\
{}[O\,{\sc iii}] 5006.84 & 5006.2 & 25.2 (6) & 21.8 (6) \\
{}[Fe\,{\sc ii}] 5111.63 & 5108.0 & 5.5 (28) & 4.7 (28) \\
{}[Fe\,{\sc ii}] 5158.81 & 5159.0 & 31.0 (4) & 23.5 (4) \\
{}[N\,{\sc i}] 5197.9 + 5200.3 & 5199.0 & 18.8 (6) & 13.8 (6) \\
{}[Fe\,{\sc ii}] 5220.1 & 5219.8 & 4.5 (19) & 3.2 (19) \\
{}[Fe\,{\sc ii}] 5261.61+5270.40 & 5265.0 & 26.6 (8) & 18.5 (8) \\
{}[Fe\,{\sc ii}] 5333.6 & 5333.9 & 10.8 (11) & 7.1 (11) \\
{}[Fe\,{\sc ii}] 5376.5 & 5376.4 & 8.5 (12) & 5.4 (12) \\
{}[Fe\,{\sc ii}] 5412.64 & 5411.6 & 5.5 (21) & 3.4 (21) \\
{}[Fe\,{\sc ii}] 5433.1 & 5433.1 & 3.0 (35) & 1.8 (35) \\
{}[Fe\,{\sc ii}] 5527.33 & 5526.6 & 13.0 (10) & 7.4 (10) \\
{}[O\,{\sc i}] 5577.34 & 5577.8 & 14.8 (8) & 8.2 (8) \\
{}[N\,{\sc ii}] 5754.64 & 5753.2 & 10.6 (12) & 5.3 (12) \\
\tiny{(+[Fe\,{\sc ii}] 5752.5+5755.6?)} & & & \\
He\,{\sc i} 5875.64 & 5876.0 & 18.9 (6) & 8.7 (6) \\
{}[O\,{\sc i}] 6300.30 & 6301.0 & 450.8 (1) & 166.3 (2) \\
{}[O\,{\sc i}] 6363.78 & 6364.2 & 154.5 (1) & 55.2 (3) \\
{}[N\,{\sc ii}] 6548.03 & 6547.9 & 36.7 (4) & 11.9 (5) \\
H$\alpha$ 6562.82 & 6563.3 & 882.8 (0.4) & 285.1 (2) \\
{}[N\,{\sc ii}] 6583.41 & 6583.4 & 107.3 (2) & 34.3 (3) \\
{}[S\,{\sc ii}] 6716.47 & 6717.1 & 79.2 (2) & 23.7 (3) \\
{}[S\,{\sc ii}] 6730.85 & 6731.5 & 164.6 (1) & 48.9 (3) \\
He\,{\sc i} 7065.28 & 7064.2 & 10.7 (17) & 2.7 (17) \\
{}[Fe\,{\sc ii}] 7155.14 & 7154.9 & 72.6 (3) & 17.4 (4) \\
{}[Fe\,{\sc ii}] 7172.0 & 7171.2 & 19.1 (8) & 4.5 (9) \\
{}[Fe\,{\sc ii}] 7290.97 & 7291.0 & 30.2 (6) & 6.8 (6) \\
{}[O\,{\sc ii}] 7319.92 & 7319.1 & 74.4 (4) & 16.4 (5) \\
{}[O\,{\sc ii}] 7330.19 & 7328.6 & 74.3 (4) & 16.3 (5) \\
{}[Ni\,{\sc \,{\sc ii}}] 7377.83 & 7377.8 & 36.9 (9) & 7.9 (9) \\
{}[Fe\,{\sc ii}] 7388.16 & 7388.2 & 16.3 (18) & 3.5 (18) \\
{}[Ni\,{\sc ii}] 7411.61 & 7410.6 & 11.4 (26) & 2.4 (26) \\
{}[Fe\,{\sc ii}] 7452.54 & 7452.2 & 26.4 (6) & 5.4 (7) \\
Ca\,{\sc ii}  8498.0 + Pa16 8502.5 & 8503.9 & 56.6 (5) & 7.2 (6) \\
Ca\,{\sc ii} 8542.1 + Pa15 8545.4 & 8547.8 & 63.5 (4) & 7.9 (6) \\
Pa14 8598.4 & 8602.3 & 4.9 (43) & 0.6 (43) \\
{}[Fe\,{\sc ii}] 8617 & 8619.6 & 63.3 (5) & 7.7 (6) \\
Ca\,{\sc ii} 8662.1 + Pa13 8665.0 & 8667.7 & 54.6 (6) & 6.5 (7) \\
\hline
\end{tabular}
\end{table}

\begin{figure*}[!t]
   \begin{center}
   \includegraphics[width=0.6\linewidth,angle=-90]{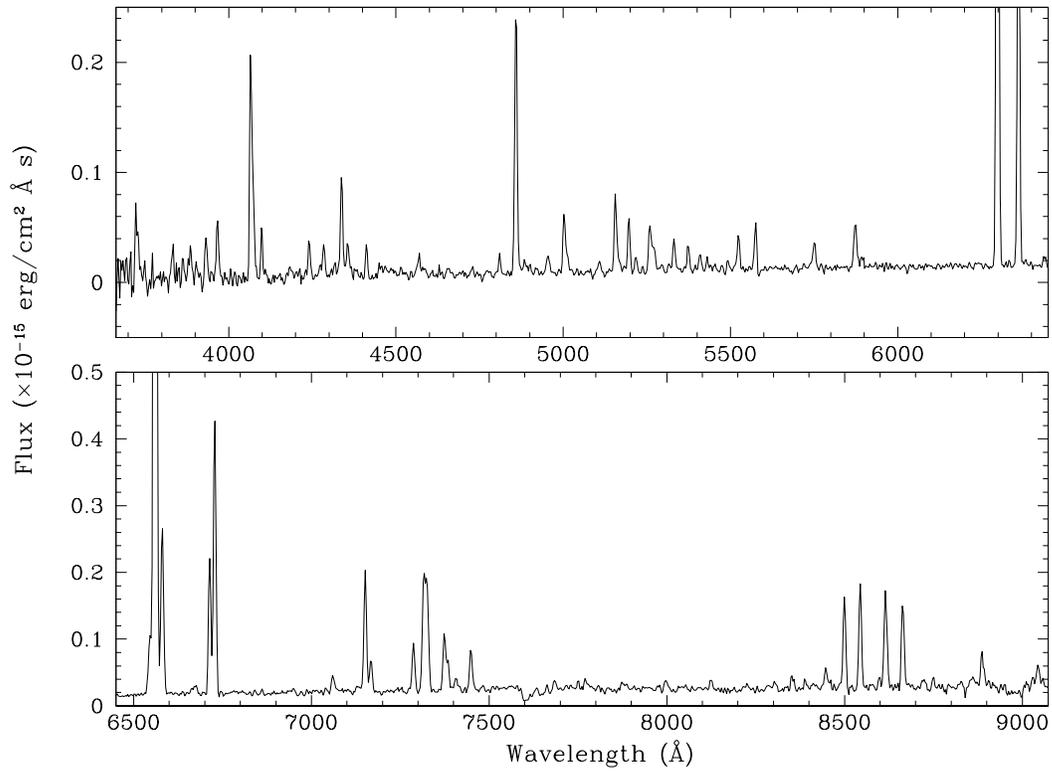}
      \caption{\label{fig:8}NOT+ALFOSC spectrum of PN1.}
      \end{center}
\end{figure*}

\begin{table}
\caption{\label{tab:4} As in Table~\ref{tab:6} for the INT+IDS spectrum of PN2.}
 \centering
 \begin{tabular}{@{}lllll@{}}
 \hline \hline
Line, $\lambda$ & $\lambda$~Obs.& F (Obs.) & F (Dr.)\\
 \hline
H7 3970.07 & 3968.9 &  12.9 (29) & 41.1 (29) \\
{}[S\,{\sc ii}] 4068.60 & 4068.0 &    49.4 (7) & 139.3 (7) \\
{}[S\,{\sc ii}] 4076.35 & 4075.9 &    13.9 (25) & 38.7 (25) \\
H$\delta$ 4101.74 & 4102.0 & 	      13.9 (25) & 37.7 (25) \\
H$\gamma$ 4340.47 & 4339.9 & 	      31.0 (8) & 61.1 (8) \\
{}[Fe\,{\sc ii}] 4416.27 & 4416.6 &   10.8 (32) & 19.1 (32) \\
H$\beta$ 4861.33 & 4861.0 & 	      100.0 (3) & 100.0 (3) \\
{}[O\,{\sc iii}] 5006.84 & 5005.9 &   16.3 (12) & 13.9 (12) \\
{}[Fe\,{\sc ii}] 5158.81 & 5158.0 &   15.1 (13) & 11.0 (13) \\
{}[N\,{\sc i}] 5197.9 + 5200.3 & 5198.1 & 4.9 (44) & 3.5 (44) \\
{}[Fe\,{\sc ii}] 5261.61 & 5260.1 &   8.0 (25) & 5.3 (25) \\
{}[Fe\,{\sc ii}] 5333.60 & 5333.0 &   4.7 (37) & 2.9 (37) \\
{}[Fe\,{\sc ii}] 5376.50 & 5376.1 &   4.2 (36) & 2.5 (36) \\
{}[Fe\,{\sc ii}] 5527.33 & 5526.5 &   4.9 (48) & 2.6 (48) \\
{}[O\,{\sc i}] 5577.34 & 5577.1 &     14.1 (17) & 7.2 (17) \\
{}[N\,{\sc ii}] 5754.64 & 5753.1 &    8.9 (26) & 4.0 (26) \\
He\,{\sc i} 5875.64 & 5874.3 &        23.6 (11) & 9.8 (11) \\
{}[O\,{\sc i}] 6300.30 & 6299.2 &     299.6 (1) & 96.0 (3) \\
{}[O\,{\sc i}] 6363.78 & 6362.7 &     97.2 (2) & 30.0 (3) \\
{}[N\,{\sc ii}] 6548.03 & 6547.8 &    17.9 (15) & 5.0 (15) \\
H$\alpha$ 6562.82 & 6562.3 & 	      1021.1 (0.4) & 281.1 (3) \\
{}[N\,{\sc ii}] 6583.41 & 6581.9 &    35.7 (5) & 9.7 (6) \\
{}[S\,{\sc ii}] 6716.47 & 6715.2 &    25.0 (7) & 6.3 (8) \\
{}[S\,{\sc ii}] 6730.85 & 6729.5 &    56.2 (3) & 14.1 (4) \\
He\,{\sc i} 7065.30 & 7063.7 &        6.7 (30) & 1.4 (30) \\
{}[Fe\,{\sc ii}] 7155.14 & 7153.8 &   22.1 (8) & 4.3 (9) \\
{}[Fe\,{\sc ii}] 7172.0 & 7170.2 &   4.7 (40) & 0.9 (40) \\
{}[Fe\,{\sc ii}] 7223.83? & 7222.4 &   21.1 (9) & 4.0 (10) \\
{}[O\,{\sc ii}] 7319.70 & 7318.4 &    44.9 (8) & 8.0 (9) \\
{}[O\,{\sc ii}] 7330.20 & 7327.9 &    40.3 (9) & 7.1 (9) \\
{}[Ni\,{\sc ii}] 7377.83 & 7376.1 &   9.4 (23) & 1.6 (24) \\
{}[Fe\,{\sc ii}] 7388.16 & 7385.6 &   6.2 (35) & 1.1 (36) \\
{}[Fe\,{\sc ii}] 7452.54 & 7450.8 &   5.9 (34) & 1.0 (35) \\
O\,{\sc i} 8446.48 & 8446.9 & 	      30.3 (22) & 2.9 (22) \\
Ca\,{\sc ii} 8498.0 + Pa16 8502.5 & 8498.7 & 59.9 (11) & 5.7 (13) \\
Ca\,{\sc ii} 8542.1 + Pa15 8545.4 & 8543.7 & 65.1 (10) & 6.0 (11) \\
Ca\,{\sc ii} 8662.1 + Pa13 8665.0 & 8664.4 & 77.9 (24) & 6.9 (25) \\
\hline
\end{tabular}
\end{table}

\begin{figure*}[!t]
   \begin{center}
   \includegraphics[width=0.6\linewidth,angle=-90]{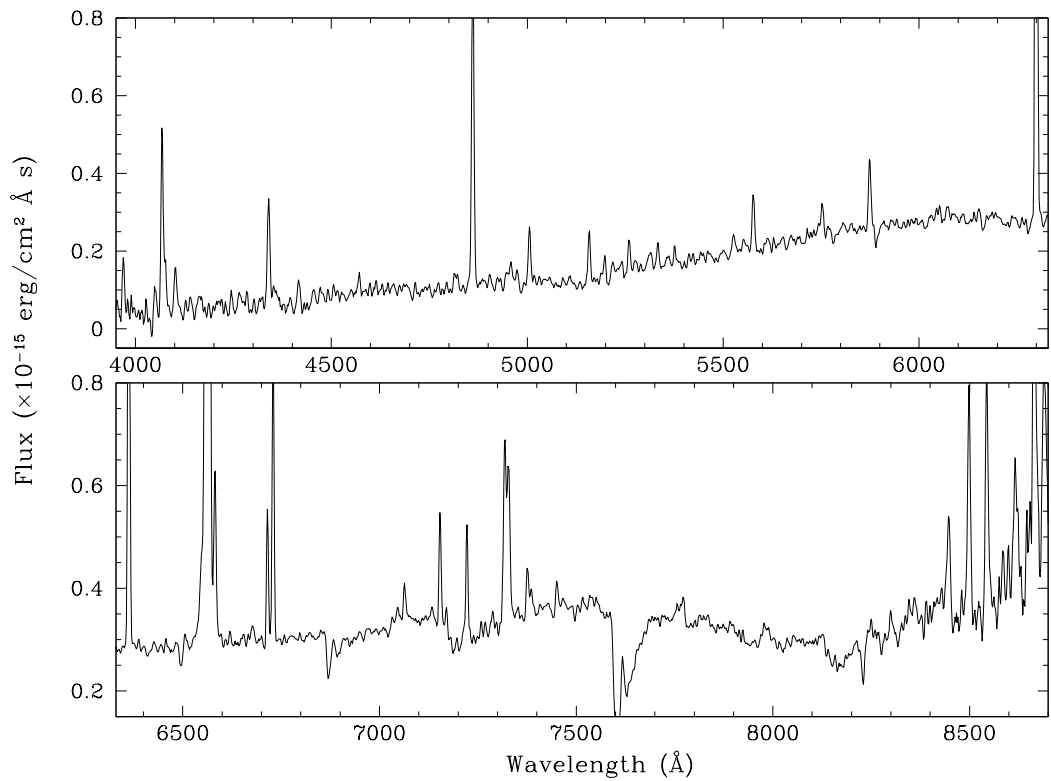}
      \caption{\label{fig:12} The INT+IDS spectrum of PN2.}
      \end{center}
\end{figure*}

\begin{figure*}[!t]
   \begin{center}
   \includegraphics[width=0.6\linewidth,angle=-90]{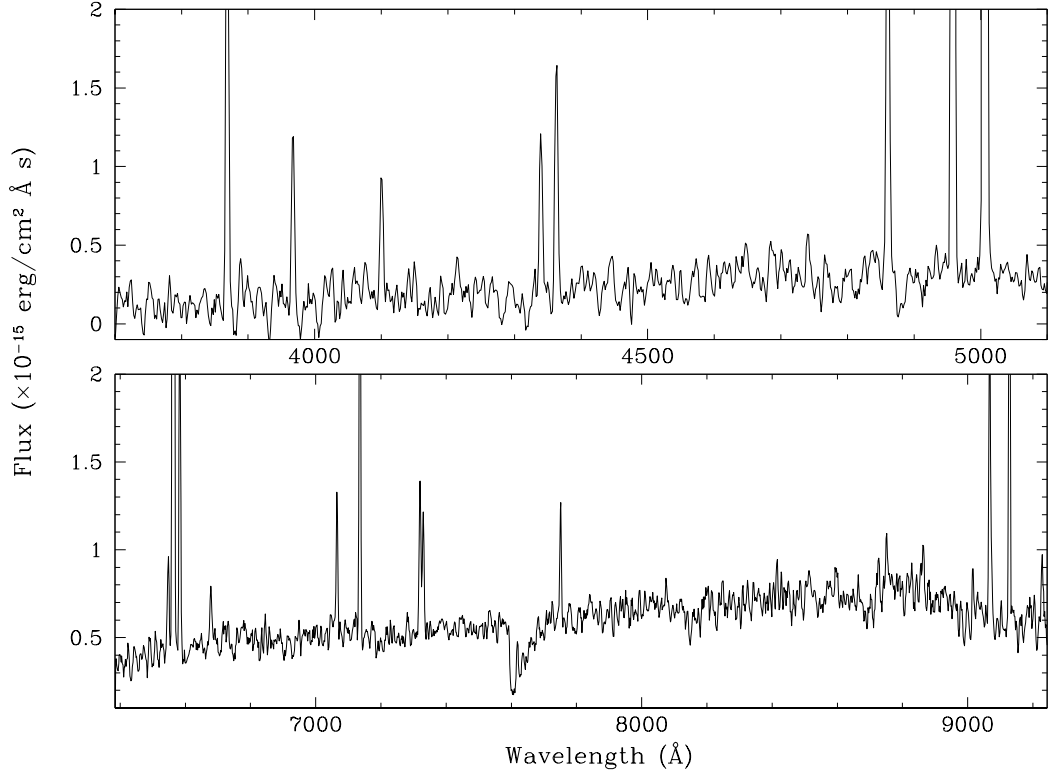}
      \caption{\label{fig:14} The WHT+ISIS spectrum of PN3.}
      \end{center}
\end{figure*}

\begin{table}
\caption{\label{tab:5} As in Table~\ref{tab:6} for the WHT+ISIS spectrum of PN3. }
 \centering
 \begin{tabular}{@{}lllll@{}}
 \hline \hline
Line, $\lambda$ & $\lambda$~Obs. & F (Obs.) & F (Dr.)\\
 \hline
{}[Ne\,{\sc iii}] 3868.75 & 3868.6 & 99.7 (3) & 197.8 (5) \\
{}[Ne\,{\sc iii}] 3967.46 & 3967.2 & 37.3 (14) & 69.8 (14) \\
H$\delta$ 4101.74 & 4100.7 & 	     23.4 (16) & 40.1 (16) \\
H$\gamma$ 4340.47 & 4339.5 & 	     36.5 (7) & 52.8 (8) \\
{}[O\,{\sc iii}] 4363.21 & 4362.8 &  50.7 (5) & 72.1 (5) \\
H$\beta$ 4861.33 & 4860.7 & 	     100.0 (4) & 100.0 (4) \\
{}[O\,{\sc iii}] 4958.91 & 4958.2 &  414.4 (1) & 390.1 (1) \\
{}[O\,{\sc iii}] 5006.84 & 5006.2 &  1166.1 (0) & 1067.0 (1 \\
{}[N\,{\sc ii}] 6548.03 & 6547.7 &   17.1 (13) & 8.5 (14) \\
H$\alpha$ 6562.82 & 6562.8 & 	     566.4 (1) & 281.1 (4) \\
{}[N\,{\sc ii}] 6583.41 & 6583.0 &   64.8 (3) & 31.9 (5) \\
He\,{\sc i} 6678.15 & 6678.2 &       12.6 (16) & 6.0 (16) \\
He\,{\sc i} 7065.30 & 7065.3 &       28.0 (9) & 11.9 (10) \\
{}[Ar\,{\sc iii}] 7135.78 & 7135.6 & 78.5 (3) & 32.5 (6) \\
{}[O\,{\sc ii}] 7319.70 & 7320.0 &   34.9 (8) & 13.7 (9) \\
{}[O\,{\sc ii}] 7330.20 & 7329.8 &   26.0 (7) & 10.1 (9) \\
{}[Ar\,{\sc iii}] 7751.10 & 7751.1 & 23.1 (10) & 7.9 (12) \\
P40 8245.64 & 8248.4 &		     4.9 (59) & 1.5 (59) \\
P19 8413.32 & 8414.4 &		     6.6 (27) & 1.9 (28) \\
P12 8750.47 & 8750.8 &		     12.1 (32) & 3.2 (33) \\
P11 8862.79 & 8863.6 &		     8.8 (30) & 2.3 (31) \\
P10 9014.91 & 9015.2 &		     9.1 (28) & 2.3 (29) \\
{}[S\,{\sc iii}] 9068.90 & 9067.2 &  57.1 (5) & 14.1 (9) \\
P9 9229.01 & 9227.8 &		     13.1 (32) & 3.2 (33) \\
\hline
\end{tabular}
\end{table}

\begin{figure}
   \begin{center}
   \includegraphics[width=0.8\linewidth,angle=0]{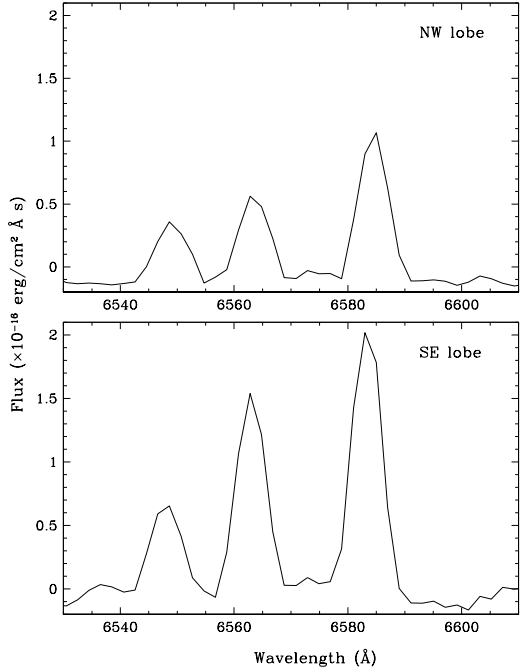}
      \caption{\label{fig:6} Part of the SPM spectrum of PN4 at P.A.=157$^{\circ}$ showing the H$\alpha$ and [N\,{\sc ii}] $\lambda\lambda6548,6583$ lines for both the NW and the SE lobe.}
      \end{center}
\end{figure}

\begin{figure*}
   \begin{center}
   \includegraphics[width=0.6\linewidth,angle=-90]{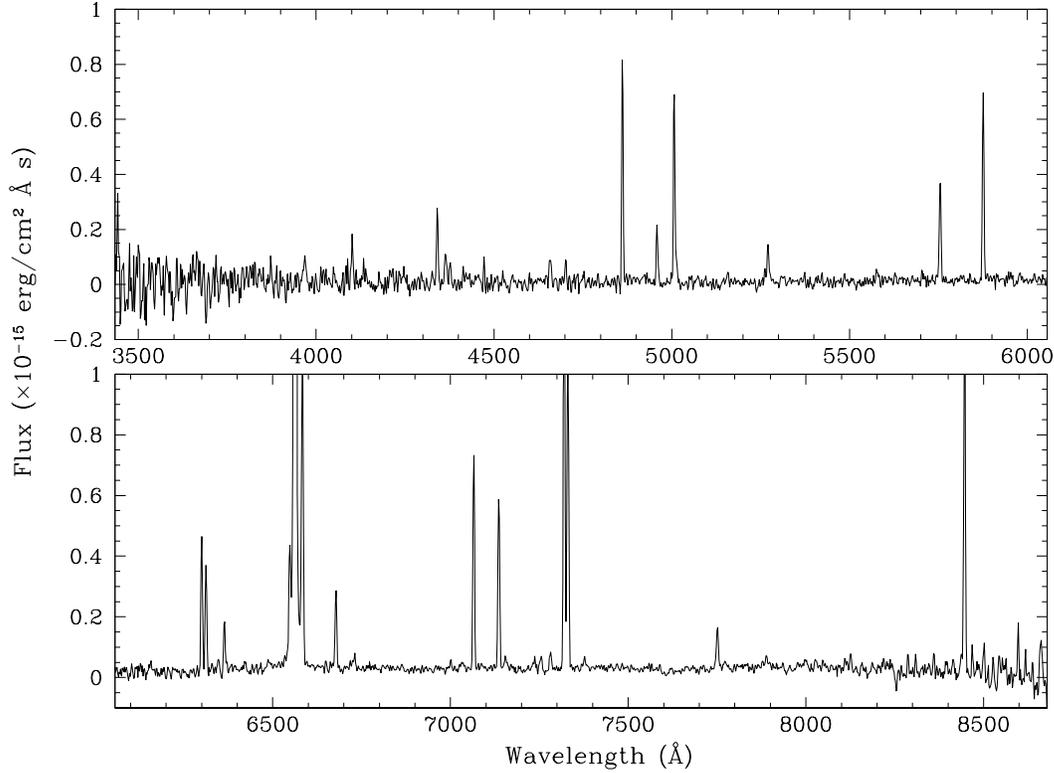}
      \caption{\label{fig:7} The INT+IDS spectrum of the nucleus of PN4.}
      \end{center}
\end{figure*}

\begin{table}
\caption{\label{tab:2} As in Table~\ref{tab:6} for the INT+IDS spectrum of PN4.}
 \centering
 \begin{tabular}{@{}lll@{}}
 \hline \hline
Line, $\lambda$ & $\lambda$ Obs. & F (Obs.)\\
 \hline
{}[Ne\,{\sc iii}] 3967.46 & 3967.3 & 20.8 (25) \\
H\,$\delta$ 4101.74 & 4101.6 	   & 22.6 (25) \\
H\,$\gamma$ 4340.47 & 4340.6       & 36.0 (10) \\
{}[O\,{\sc iii}] 4363.21 & 4364.2  & 18.1 (21) \\
{}[Fe\,{\sc ii}] 4373.7 & 4377.1   & 11.7 (26) \\
He\,{\sc i} 4471.09 & 4471.9       & 8.5 (29) \\
{}[Fe\,{\sc iii}] 4658.10 & 4658.4 & 15.1 (18) \\
{}[Fe\,{\sc iii}] 4701.62 & 4701.6 & 11.3 (31) \\
H\,$\beta$ 4861.33 & 4861.5        & 100.0 (2) \\
{}[O\,{\sc iii}] 4958.91 & 4958.7  & 29.7 (7) \\
{}[O\,{\sc iii}] 5006.84 & 5006.9  & 103.0 (2) \\
{}[Fe\,{\sc ii}] 5158.81 & 5157.6  & 6.9 (40) \\
{}[Fe\,{\sc ii}] 5261.61 & 5262.2   & 5.1 (23) \\
{}[Fe\,{\sc iii}] 5270.40 + [Fe\,{\sc ii}] 5274.9 & 5270.6 & 22.3 (8) \\
{}[N\,{\sc ii}] 5754.64 & 5754.6   & 55.0 (4) \\
He\,{\sc i} 5875.64 & 5875.7       & 94.3 (2) \\
{}[O\,{\sc i}] 6300.30 & 6300.4    & 65.6 (2) \\
{}[S\,{\sc iii}] 6312.10 & 6312.1  & 50.9 (4) \\
{}[O\,{\sc i}] 6363.78 & 6363.8    & 21.2 (7) \\
{}[N\,{\sc ii}] 6548.03 & 6549.6   & 108.3 (2) \\
H\,$\alpha$ 6562.82 & 6563.0       & 5412.9 (0.1) \\
{}[N\,{\sc ii}] 6583.41 & 6583.3   & 166.4 (1) \\
He\,{\sc i} 6678.15 & 6678.0       & 36.3 (6) \\
{}[S\,{\sc ii}] 6716.47 & 6719.2   & 3.7 (70) \\
{}[S\,{\sc ii}] 6730.85 & 6730.9   & 3.4 (55) \\
He\,{\sc i} 7065.28 & 7065.3       & 106.1 (1) \\
{}[Ar\,{\sc iii}] 7135.78 & 7135.8 & 87.0 (2) \\
{}[Fe\,{\sc ii}] 7155.14 & 7155.1  & 8.2 (26) \\
C\,{\sc ii} 7236.42 & 7235.6 & 8.0 (25) \\
O\,{\sc i} 7254.4 & 7254.6 & 8.0 (22) \\
He\,{\sc i} 7281.35 & 7281.4       & 10.4 (18) \\
{}[O\,{\sc ii}] 7319.4 & 7319.9    & 212.9 (0.9) \\
{}[O\,{\sc ii}] 7330.3 & 7330.4    & 165.5 (1) \\
{}[Ni\,{\sc ii}] 7377.83 & 7377.5  & 5.9 (30) \\
{}[Ar\,{\sc iii}] 7751.10 & 7751.1 & 23.5 (8) \\
P30 8286.43 & 8286.7               & 7.0 (40) \\
P27 8306.11 & 8308.5               & 5.4 (44) \\
P22 8359.00 & 8359.7               & 7.1 (39) \\
P18 8437.96 & 8437.6               & 22.8 (39) \\
O\,{\sc i} 8446.48 & 8446.6        & 160.9 (5) \\
P17 8467.25 & 8468.3               & 11.3 (30) \\
P16 8502.48 & 8500.9               & 14.5 (26) \\
P14 8598.39 & 8597.4               & 16.0 (33) \\
{}[Fe\,{\sc ii}] 8617 & 8618.0  & 7.9 (51) \\
P13 8665.02 & 8661.5               & 30.2 (29) \\
\hline
\end{tabular}
\end{table}

\begin{table}
\caption{\label{tab:3} As in Table~\ref{tab:6} for the INT+IDS spectrum of PN5 at $P.A.$= 295$^{\circ}$ and at three different parts of the nebula.}
 \centering
 \begin{tabular}{@{}lllll@{}}
 \hline \hline
Line, $\lambda$ & Part & $\lambda$~Obs. & F (Obs.) & F (Dr.)\\
 \hline
He\,{\sc ii} 4685.7 & NW & 4683.4 & 108.0 (16) & 125.2 (16) \\
 & cent & 4685.0 & 118.2 (41) & 130.3 (42) \\
 & SE & 4686.6 & 78.4 (16) & 89.5 (16) \\
& & & & \\
H$\beta$ 4861.33 & NW & 4860.4 & 100.0 (14) & 100.0 (14) \\
 & cent & 4861.4 & 100.0 (35) & 100.0 (35) \\
 & SE & 4861.1 & 100.0 (16) & 100.0 (16) \\
& & & & \\
{}[O\,{\sc iii}] 4958.91 & NW & 4958.7 & 417.1 (4) & 386.4 (4) \\
 & cent & 4958.9 & 171.8 (21) & 163.3 (21) \\
 & SE & 4958.8 &  411.5 (4) & 384.5 (4)\\
& & & & \\
{}[O\,{\sc iii}] 5006.84 & NW & 5006.9 & 1385.8 (1) & 1238.8 (2) \\
 & cent & 5006.9 & 609.3 (6) & 565.9 (7) \\
 & SE & 5006.8 & 1345.5 (1) & 1217.6 (2) \\
& & & & \\
H$\alpha$ 6562.82 & NW & 6562.8 & 690.4 (2) & 285.1 (15) \\
 & cent & 6562.9 & 510.7 (8) & 285.1 (37) \\
 & SE & 6562.9 & 626.9 (3) & 285.1 (16) \\
& & & & \\
{}[N\,{\sc ii}] 6583.41 & NW & 6583.5 & 254.9 (5) & 104.4 (15) \\
 & SE & 6583.2 & 183.6 (8) & 82.9 (18) \\
\hline
\end{tabular}
\end{table}

\begin{figure*}
   \begin{center}
   \includegraphics[width=0.5\linewidth,angle=-90]{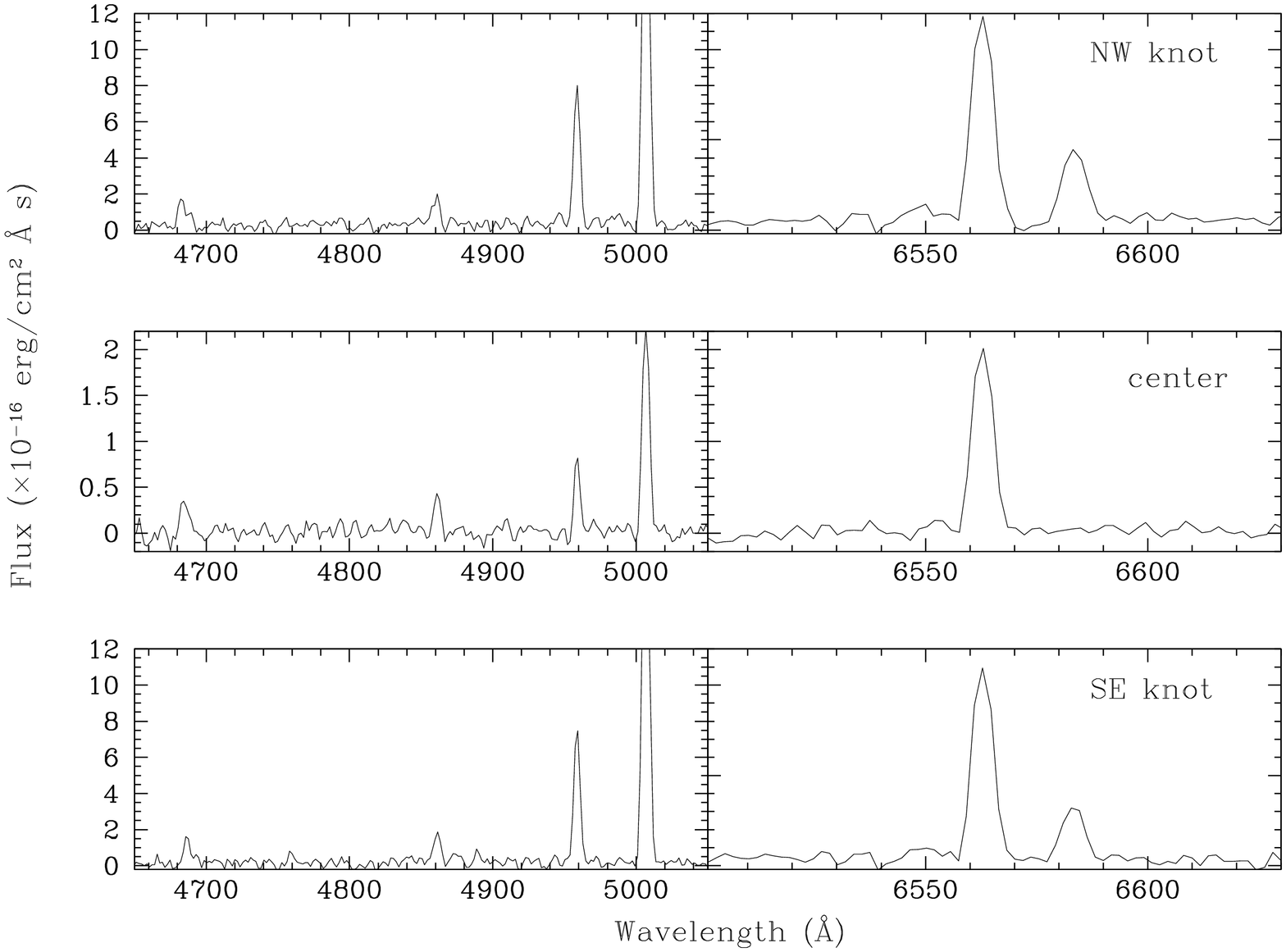}
      \caption{\label{fig:10} Parts of the INT+IDS spectrum of PN5 at P.A.=295$^{\circ}$, showing the two [O\,{\sc iii}] bright knots, and the centre of the PN. Note the different scale in the spectrum of the centre.}
      \end{center}
\end{figure*}

\subsection{Spectral features}

The spectra for PN1, PN2, and PN3, are shown in Figs.~\ref{fig:8}, \ref{fig:12}, and \ref{fig:14}, respectively, and the measured emission lines are listed in Tables~\ref{tab:6}, \ref{tab:4} and \ref{tab:5}, respectively.

In addition to typical low and intermediate excitation PN lines, the spectrum of PN1 shows many [Fe\,{\sc ii}] lines as well as the Ca\,{\sc ii} infrared triplet in emission.  These lines are not usual in normal, evolved PNe but are present in a large
variety of astronomical objects, including
interacting binary systems (such as cataclysmic variables, symbiotic
stars, slow novae, and VV Cephei stars), young active stars (some Be
stars, T-Tauri stars, and other pre-main-sequence objects), active
galactic nuclei, and proto-PNe \citep[see][and references
therein]{corradi95,rodriguez01}. The continuum of PN1 is very weak with no absorption bands (the band at $\sim$ 7600~\AA, visible also in PN2, PN3, and PN4 is of atmospheric origin).

The spectrum of PN2 resembles that of PN1 in being typical of low to intermediate excitation PNe and showing [Fe\,{\sc ii}] and Ca\,{\sc ii} triplet in emission. Unlike PN1, it shows a clear continuum. The steep increase in the continuum at the red end of the spectrum beyond 8200\AA~is unlikely to be real, but due to poor calibration data at these wavelengths.

The spectrum of PN3 differs from the two previous ones in that iron and calcium are not seen in emission, while several intermediate excitation lines, typical of PNe, are present: [Ne\,{\sc iii}], [Ar\,{\sc iii}] and [S\,{\sc iii}]. A clear, featureless continuum is present.

The SPM spectrum of PN4 allows the identification of only H$\alpha$, [N\,{\sc ii}] $\lambda$6548, and [N\,{\sc ii}] $\lambda$6584 lines in the lobes and in the nucleus. A cut of this spectrum showing the detected lines is presented in Fig.~\ref{fig:6}. The I($\lambda$6548+$\lambda$6583)/I(H$\alpha$) line ratio in the lobes is notably high: 2.6 $\pm$ 0.4 for the NW lobe and 1.8 $\pm$ 0.2 in the SE one. High [N\,{\sc ii}]/H$\alpha$ ratios can be produced by shocks or by a high nitrogen abundance: given the faintness of the [S\,{\sc ii}] lines, a high nitrogen abundance is more likely.  This is common among bipolar PNe \citep{perinotto98}. From the [N\,{\sc ii}] $\lambda$6583 line, we measure a difference in radial velocity of $47 \pm 13$ kms$^{-1}$ between the lobes. Unless the major axis of the nebula lies close to the plane of the sky, this small value indicates a moderate expansion velocity for the bipolar lobes.
The IDS spectrum of the core of PN4 is shown in Fig.~\ref{fig:7}; fluxes are given in Table~\ref{tab:2}. The spectrum is characterised by lines typical of low- to intermediate-excitation PNe (including [Ne\,{\sc iii}] and [Ar\,{\sc iii}]) and a very weak continuum without apparent absorption bands. Both [Fe\,{\sc ii}] and [Fe\,{\sc iii}] lines are seen in emission. The H$\alpha$ line is 54 times stronger than H$\beta$, a clear sign of self-absorption at the core of PN4.

When studying the H$\alpha$ image of the W08 object (marked by an arrow in the [N\,{\sc ii}] image in Fig.~\ref{fig:5}), we found an extended PN, PN5, on which the image of the catalogued star is projected. The INT+IDS spectrum of PN5 is presented in Fig.~\ref{fig:10} and the line fluxes in Table~\ref{tab:3}. The object is very faint and its spectrum shows only a few lines with no continuum; it seems to correspond to a normal, evolved PN.

\subsection{Extinction}\label{ext}

Compact PNe can attain very high densities, and the presence of [Fe\,{\sc ii}] and Ca\,{\sc ii} lines together with anomalous line ratios (as will be shown in the next section) in the new objects warns us to study possible optical depth effects in the Balmer lines \citep{drake80}, which can complicate or even preclude estimating the interstellar extinction. \cite{phillips06} has reviewed the effect of self-absorption on the Balmer line ratios for PNe. Using his $\mbox{H}\alpha/\mbox{H}\beta - \mbox{H}\gamma/\mbox{H}\beta$ diagram, we find that the core of PN4 clearly suffers from self-absorption. Although the electron density in its bipolar lobes should be lower, this is of little help in that the H$\beta$ line was not detected in the SPM spectrum of either lobe. From this non-detection, a 3-$\sigma$ lower limit can be calculated: $I(\mbox{H}\alpha)/I(\mbox{H}\beta) > 6.7$, giving $c > 1.3$ or $A_V > 2.7$ mag. However, in section 7.1 we argue that the interstellar extinction for PN4 is likely to be
 $A_V\approx 3.2$ $\pm$ 0.5 mag, and we use this tentative value in the following when needed. 

On the other hand, PN1, PN2, PN3, and PN5 are not self-absorbed, enabling the reddening constants to be calculated. Using the reddening law of \cite{cardelli89} for $R_V$ = 3.1, we obtain $c$ = 1.65 $\pm$ 0.04, 1.88 $\pm$ 0.04, and 1.02 $\pm$ 0.06, for PN1, PN2, and PN3, respectively. For PN5, the extinction can be calculated from the $I(H\alpha)/I(H\beta)$ ratio separately for three zones of the nebula: the NW knot, the SE knot and the centre, giving $c=1.3\pm0.2$, $0.9\pm0.5$, $1.2 \pm 0.2$, respectively. No significant extinction variation is thus found to within the errors, and the mean reddening constant derived from the summed spectrum $c=1.19\pm0.15$ is adopted.

\subsection{Physical conditions}

To estimate the electron density ($N_e$) and temperature ($T_e$) of the objects under study, we considered all sensitive line ratios measured from their optical spectra. In the following, we shall use the abbreviations: [S\,{\sc ii}] density-sensitive line-ratio: $I(\lambda6716)/I(\lambda6731)$ (S2$_{N_e}$), [S\,{\sc ii}] temperature-sensitive line-ratio ($I(\lambda6716+ \lambda6731)/I(\lambda4068+ \lambda4076)$ (S2$_{T_e}$), [O\,{\sc iii}] temperature-sensitive line-ratio $I(\lambda4959+\lambda5007)/I(\lambda4363)$ (O3), [N\,{\sc ii}] temperature-sensitive line-ratio $I(\lambda6548+ \lambda6583)/I(\lambda5755)$ (N2), [O\,{\sc ii}] temperature-sensitive line-ratio $I(\lambda3726+ \lambda3729)/I(\lambda7320+ \lambda7330)$ (O2), [O\,{\sc i}] temperature-sensitive line-ratio $I(\lambda6300+ \lambda6363)/I(\lambda5577)$ (O1).

PN1: The S2$_{N_e}$ ratio gives a high density $N_e \ge (1.8 \pm 0.7) \times 10^4$ cm$^{-3}$. We note that this line ratio is in the high-density limit, where its density sensitivity decreases and the upper limit to the error is underestimated because of the non-Gaussian response of the density to the line ratio's variation: the true density can be much higher. Using this density, we obtain an O2 temperature $T_e = 1.2 \pm 0.2 \times 10^4$~K, whereas the O1 diagnostic gives a similar value $T_e = 1.5 \pm 0.1 \times 10^4$~K. The measured S2$_{T_e}$ line ratio has an anomalously low value indicating high density. It leads to temperatures of $T_e < 10^4$~K and requires an even higher density of $N_e > 8 \times 10^4$ cm$^{-3}$ to give physically plausible results.
Finally, the commonly used strong-line ratios O3 = 1.46 $\pm$ 0.07 and N2 = 8.75 $\pm$ 1.04 are clearly anomalously low indicating extremely high densities \citep{gurzadyan70}. We note that these line ratios should be used as lower limits in the case of PN1 because both the [O\,{\sc iii}] $\lambda$4363 and [N\,{\sc ii}] $\lambda$5755 lines may be blended with [Fe\,{\sc ii}] lines. However, the presence of these iron lines and the measured [Fe\,{\sc ii}] line intensities also indicate very high densities for the object \citep{bautista96}. Similarly, the strong infrared CaII lines around 8500{\AA} indicate very high densities from an emission region (possibly a disk) near the central star \citep{rodriguez01}.
We conclude that this object has a very high density and at least some of its line ratios suffer from density effects giving anomalous results.

PN2: From the S2$_{Ne}$ line ratio, we deduce that the density of the nebula is in the high density regime where sulphur lines cease to be density sensitive. A minimum density required to derive a reasonable physical temperature from the anomalous S2$_{T_e}$ is $N_e \sim 2 \times 10^5$ cm$^{-3}$, leading to an electron temperature of the order $10^4$~K (higher density leads to a smaller temperature). A temperature of $10^4$~K is derived from the anomalous N2 = 3.7 $\pm$ 1.0 line ratio, if a density $N_e \sim 6 \times 10^5$ cm$^{-3}$ is assumed. For this density, the O1 diagnostic gives $T_e = 1.2 \pm 0.1 \times 10^4$~K. The [Fe\,{\sc ii}] line intensities and the CaII IR triplet also once more indicate extremely high densities: the three CaII infrared lines show similar intensities and this, together with the absence of the forbidden [CaII] 7234 and 7291{\AA} lines, indicates that $N_e > 10^{10}$ cm$^{-3}$. So, PN2 is similar to PN1 in showing a high electron density (possibly even higher for PN2) and anomalous line ratios.

PN3: The O3 ratio points again to a very high density. From the dereddened line-ratio O3  = 20.2 $\pm$ 1.1, realistic electron temperatures are calculated only if an extremely high density is assumed: if $T_e$ is assumed to be in the range $10^4$ - 2 $\times$ $10^4$~K, $N_e$ ranges between 3 $\times$ $10^5$~cm$^{-3}$ and 25 $\times$ $10^5$~cm$^{-3}$.

PN4: The O3 and N2 line ratios measured from the undereddened spectrum are once more anomalously low, at respectively $7.3 \pm 1.6$ and $5.0 \pm 0.2$. The O3 line-ratio implies electron densities of the order $\sim 10^7$~cm$^{-3}$ if a normal PN temperature of $10^4$~K is adopted. Correction for reddening would increase the derived density further still.

PN5: The measured emission lines (Table 8) do not allow either the density or the temperature to be estimated for this faint object. The He II line at 4686{\AA}, only detected for this object in our sample, shows an intensity comparable to H$\beta $, indicating a very high excitation \citep[class 10 in the scheme of][]{dopita90}. The [NII] line is weak, a third of H$\alpha$, and, together with the strong [OIII] lines, confirms that PN5 is a normal, evolved, high excitation PN.

\begin{figure}
   \begin{center}
   \includegraphics[width=0.9\linewidth]{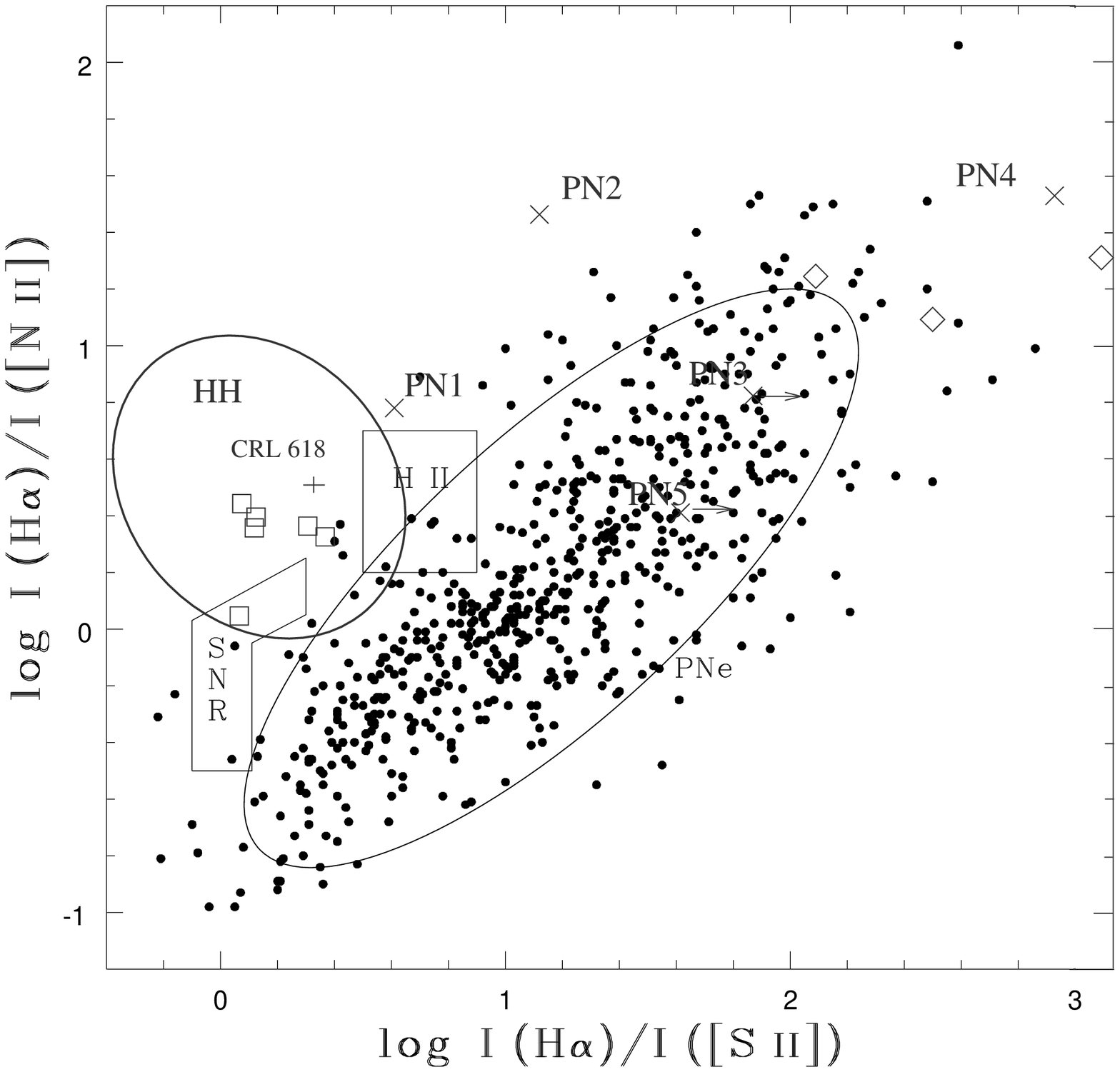}
      \caption{\label{fig:riesgodiagram} The S2N2 diagram of \cite{riesgo06} showing the locations of H\,{\sc ii} regions, SNRs and the 85\% probability ellipse of PNe. The PN sample of \cite{riesgo06} is shown as dots. We have over-plotted an ellipse surrounding the HH-objects of \citet{canto81} as well as the locations of the HH-objects of \cite{brugel81} (squares), the proto-PN CRL 618 (plus sign), the three young PNe Hb12, PM1-322 and PC11 (diamonds), together with the five objects of our sample (crosses).}
      \end{center}
\end{figure}

\begin{figure}
   \begin{center}
   \includegraphics[width=0.9\linewidth]{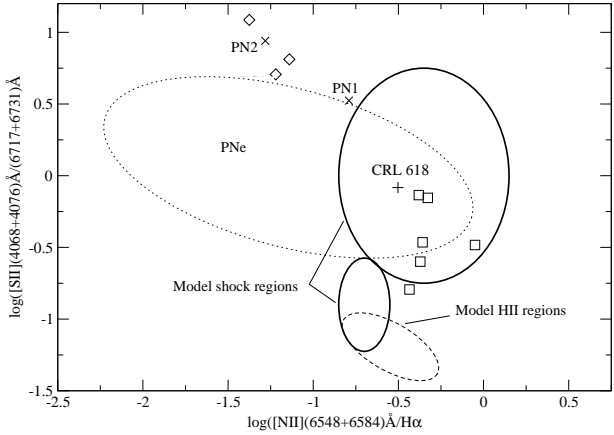}
      \caption{\label{fig:rieradiagram} Sketch of the [S\,{\sc ii}]$_{ratio}$ diagnostic diagram from \citet{riera90}. The locations of PNe,  H\,{\sc ii} regions, and shock models are shown as ellipses (see text). Symbols are as in Fig.~\ref{fig:riesgodiagram}.}
      \end{center}
\end{figure}

\begin{figure}
   \begin{center}
   \includegraphics[width=0.9\linewidth]{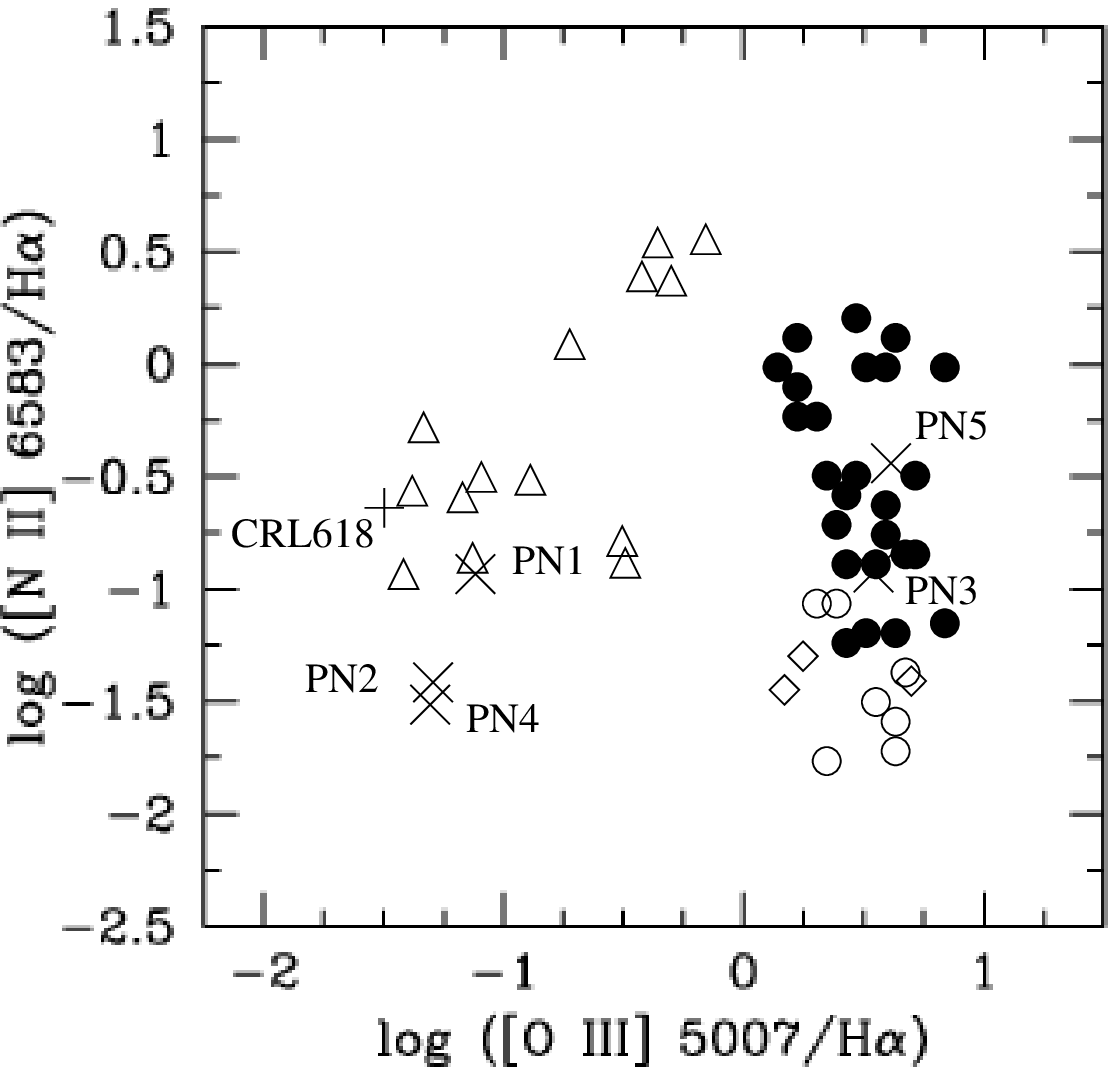}
      \caption{\label{fig:ragadiagram} The [N\,{\sc ii}]/[O\,{\sc iii}] diagram adapted from \citet{raga08}. The locations of rims and shells of PNe (empty circles), FLIERS (filled circles) and proto-PNe (triangles) are from the original paper. We have overplotted the locations of CRL618 and the young PNe as in Fig.~\ref{fig:riesgodiagram}, together with our five PNe.}
      \end{center}
\end{figure}

\section{Diagnostic diagrams}\label{s6}

To gain further insight into the nature of the candidate PNe, we have studied their location in several diagnostic diagrams.

\subsection{IPHAS and 2MASS}

Fig.~\ref{fig:1} and \ref{fig:2} show that only PN4 is located in ``zone 1'' of the IPHAS colour-colour diagram. The others (PN1, PN2, and PN3) have smaller $r^\prime - H\alpha$ colours and appear in the upper part of zone 2 (Fig.~\ref{fig:2}), where Galactic PNe are accompanied by other classes of emission-line objects. Interestingly, all these four new objects are located high up in the 2MASS diagram, in an area mostly occupied by symbiotic Miras but also by a few well-known objects classified as young PNe (M2~9, Mz~3, etc.) that have similar NIR colours to symbiotic stars and whose nature is not yet fully clear \citep{corradi08}. PN3 is at the border of the T-Tauri zone in the 2MASS diagram of Fig.~\ref{fig:2} (i.e., near the 90\% isocontour), whereas PN1, PN2, and PN4 are outside that zone.

\subsection{Infrared colour-colour diagrams}

\citet{pottasch88} studied the location of H\,{\sc ii} regions, OH/IR stars, PNe, normal late-type stars, and galaxies in the IRAS $F(12\mu$m)$/F(25\mu$m) vs. $F(25\mu$m)$/F(60\mu$m) diagram. PN1 and PN4 are IRAS sources (Section 5.1) with respective colours $F(12\mu$m)/$F(25\mu$m) $<$ 0.65; $F(25\mu$m)/$F(60\mu$m) = 0.75 and $F(12\mu$m)$/F(25\mu$m) = 0.61; $F(25\mu$ m)/$F(60\mu$ m) $>$ 0.90, which locate both objects in the zone where OH/IR stars and PNe overlap, and well away from the zones of H\,{\sc ii} regions and late-type stars. According to \citet{pottasch88}, about 20\% of the most evolved OH/IR stars have colours similar to young PNe. However, T-Tauri stars are also found in this area of the IRAS diagram \citep{magnani95}.

The combined MSX 8 to 21 $\mu m$ and 2MASS data for PN4 locate it again in the OH/IR star - PN overlap area \citep[see e.g, Fig. 2 and 3 from][]{ortiz05}, and close to the youngest PNe in the MSX evolutionary tracks of \citet{ortiz07}, where a few known proto-PNe (but also suspected symbiotic stars) appear near PN4. For PN1 and PN2, there are only MSX 8.28 $\mu m$ data (Section 5.1). This, when combined with 2MASS data, locates both objects (and PN4 also) high up in the near/mid infrared diagrams of \citet{lumsden02} for ``8$\mu m$ only'' sources (their Fig. 7) in a zone populated by very red stars (or heavily obscured sources, unlike our new discoveries - see section 5.3) that at the same time show moderate mid-infrared excesses. Although a precise classification for PN1 and PN2 is not possible based on this infrared diagram alone, according to \citet{lumsden02} it would be relatively unlikely that they are low-mass pre-main-sequence stars, or more massive (Be-type) young stars.

\subsection{Optical line diagnostic diagrams}

There are several diagnostic diagrams in the literature that use ratios of strong optical lines to identify different classes of ionized objects (e.g., PNe, H\,{\sc ii} regions, supernova remnants (SNR), and Herbig-Haro (HH) objects). At the same time, these provide constraints on the excitation mechanism at work (photoinization, shocks, or both). To study our new objects in the relevant diagrams, a first important datum is that all four compact sources, PN1, PN2, PN3, and PN4, show anomalous (very low) O3 and/or N2 ratios, indicative of very high densities, as discussed in Section 5.4. Therefore, we start our analysis by locating the sources in a diagram from \citet{gutierrez-moreno95}, who plotted the [O\,{\sc iii}] $\lambda$4363/H$\gamma$ vs. [O\,{\sc iii}] $\lambda$5007/H$\beta$ line ratios, in order to separate ionized objects of very different densities, e.g., symbiotic stars, young PNe and normal, evolved PNe.  In this diagram, PN3 is found in the area of high density young PNe, while PN1 and PN4 are in the zone characterized by the highest densities (zone C in Fig. 2 from \citet{gutierrez-moreno95}, where most symbiotic stars are located). Therefore, PN1 and PN4 could be symbiotic stars according to this diagram, or, to be more precise, they both show line ratios indicative of densities as high as those found in symbiotic stars. \citet{pereira95} showed that several young, dense PNe are also located in the zone C in this diagram. For PN2, the [O\,{\sc iii}] $\lambda$4363 line is below our detection limit and so cannot be located in the diagram. However, its N2 ratio, and other indicators discussed in Section 5.4, imply also extremely high densities for the object.

A second relevant datum is that both PN1 and PN2 show very strong [S\,{\sc ii}]$\lambda\lambda$4068,4076{\AA} emission (Section 5.2), rather unusual for evolved PNe, but typical of HH objects. Although well studied proto-PN like CRL618  \citep{westbrook75} and young PNe like Hb12, PM1-322 or PC11 \citep{aller83,pereira95,gutierrez-moreno87} show strong 4068,4076{\AA} lines also, we explored further the possibility that PN1 and PN2 are HH objects. Therefore, in the following we locate our five objects, together with CRL618, the three above-mentioned young PNe, and a sample of HH with accurate line fluxes from \citet{brugel81}, in several relevant diagrams involving the [S\,{\sc ii}] and other strong nebular lines.

\citet{riesgo06} presented a revised version of the \citet{sabbadin77} log(H$\alpha$/[N\,{\sc ii}]) vs. log(H$\alpha$/[S\,{\sc ii}]) (S2N2) diagram, which has been commonly used to separate PNe, H\,{\sc ii} regions and SNRs. \cite{canto81} added $\sim$50 HH objects to the diagram. In Fig.~\ref{fig:riesgodiagram}, we have adapted this S2N2 diagram, finding that the HH objects from \citet{brugel81} are, as expected, located in Cant\'o's HH zone, while CRL618 lies in between the HH objects and PN1 and PN2. Both PN1 and PN2, together with PN4, are located on the outskirts of the zone occupied by PNe; PN2 and PN4 are well separated from the H\,{\sc ii} region and shock-excited zones, while PN1 is near these zones. PN4 shares an extreme position in the diagram with the three young PNe. For PN3 and PN5, there are no sulphur-line detections, and 3-$\sigma$ upper limits were calculated and used instead. At the calculated limits, PN3, and PN5 are located within the 85\% probability ellipse of PNe. Looking only at the position of well-known objects in Fig.~\ref{fig:riesgodiagram}, it appears that a PN would start its evolution at the shock (proto-PN and HH) zone, move later towards the upper right corner (the young-PNe area) when photoionization already dominates, and finally shift towards the main PNe zone up to locations defined by the excitation and chemical abundances of each particular object. In this context, PN1, PN2 and PN4 could be examples of very young PNe at different evolutionary stages.

Fig.~\ref{fig:rieradiagram} shows the only diagram in the literature, to our knowledge, that uses the four [S\,{\sc ii}] 4068, 4076, 6717, and 6731{\AA} lines to study a variety of emission-line objects. This figure is a sketch adapted from \cite{riera90} and shows the approximate location of photoionised (model) H\,{\sc ii} regions (dashed ellipse), observed PNe (dotted ellipse), and the predictions of simple shock models of \citet{hartigan87} (solid ellipses; for shock velocities 20-400 kms$^{-1}$ and pre-shocked densities of 100 and 1000 cm$^{-3}$). We note that the young PNe represented in Fig.~\ref{fig:rieradiagram} have very high [S\,{\sc ii}] ratios, and are located high in the diagram, above the ellipse containing normal PNe, and close to PN1 and PN2. All these objects are clearly separated from the HH objects of \citet{brugel81}. CRL618 is located well inside the shock zone, as expected for a proto-PN nebula, and again in an intermediate position between HH objects and PN1 and PN2. If there is also an evolutionary interpretation of this diagram, it would imply the same scenario for PN1 and PN2 as the S2N2 diagram discussed above.

Finally, \cite{raga08} studied the empirical locations of proto-PNe, shells/rims of normal PNe, and low-ionization microstructures (FLIERS) in several diagnostic diagrams, and compared them with results of numerical simulations of cloudlets moving at high speed in a photoionized environment. Fig.~\ref{fig:ragadiagram} show the results for the only diagram they considered where we have the necessary line measures for all our five objects. Following \cite{raga08}, we have plotted in that figure the loci of shells and rims of PNe (open circles), of FLIERS (filled circles) and of proto-PNe (triangles), together with our PNe (crosses). We note that PN4 would move towards the right (the normal PNe area) if it suffers a local extinction in excess of the value assumed here, A$_V$=3.2 mag (but note also that an unplausible A$_V$=16 mag of local extinction would be required to reach the PNe area). It is clear that PN3 and PN5 lie in the FLIERS/PNe area, whereas PN1, PN2, and PN4
 are near the proto-PN area (the HH objects from \citet{brugel81} are also located here, as expected).  For this region, the models that fit the data are those of irradiated shocks with low ionization rates (i.e., the lowest central-star temperatures in the simulations from \cite{raga08}; c.f. their Fig. 11). Interestingly, the young PNe Hb12, PM1-322 and PC11 are all located close to the normal PNe (not the FLIERS) zone in this diagram.

\section{Discussion of the nature of the sources}\label{s7}

We summarize the information for the five newly discovered objects and discuss the probable nature of each source.

PN1: An unresolved source in H$\alpha$. The infrared properties and optical diagnostic diagrams (Sect. 6.2 and 6.3) indicate that it is not a compact H\,{\sc ii} region. PN1 has [SII] and [NII] line ratios significantly different from HH objects (c.f. Fig.~\ref{fig:rieradiagram}). A T Tauri or massive YSO is unlikely given its spectrum with little continuum and its infrared characteristics. All data at hand for PN1 are consistent with it being a dense, compact PN partly photo- and partly shock-excited.

PN2: An unresolved source in H$\alpha$ with extended faint nebulosity. Similarly to PN1, an H\,{\sc ii} region is unlikely. Neither the shape of the optical continuum of PN2 (section 5.2), nor the infrared 2MASS/MSX data, nor the presence of high excitation emission lines like [O\,{\sc iii}] $\lambda\lambda$4959+5007{\AA} are typical of T Tauri stars, although a few peculiar T Tauri stars like Th 28 do show [O\,{\sc iii}] in emission \citep{appenzeller83}. However, the emission in Th 28 (and similar stars like DG Tau) originates in associated HH-like objects \citep[c.f.][]{melnikov08} and we have shown in Section 6.3 that the diagnostic diagrams separate PN2 very clearly from the HH objects (Figures~\ref{fig:riesgodiagram} and \ref{fig:rieradiagram}). Therefore, the simplest hypothesis conforming all available data is that PN2 is similar to PN1: it is an evolved object in the transition phase between a proto-PN and a young PN, where photoionisation is already at work.

PN3: An unresolved H$\alpha$ source with faint featureless continuum and relatively high excitation lines. All the information to hand points to a young, dense PN, but probably in a later stage of evolution than PN1 and PN2.

PN4: A bipolar nebula with an unresolved high density core. As for PN3, there is little doubt that PN4 is an evolved object (c.f. IPHAS, 2MASS, and MSX diagrams; diagnostic diagrams) but it could be either a PN or a symbiotic star with an associated nebula. We discuss these two alternatives in a more general context in Section 9.

PN5: A faint extended nebula and radio source. All available data indicate that it is an evolved bipolar PN.

In summary, all five new objects can be described as PNe according to the diagnostic information presently available. PN1, PN2, PN3, and PN4 show several line-flux ratios consistent with being dense young PNe, while PN1 could be a proto/young PN partially photoionized and partially shock-ionized.

\begin{figure*}
\centering
\includegraphics[width=0.5\linewidth,angle=0]{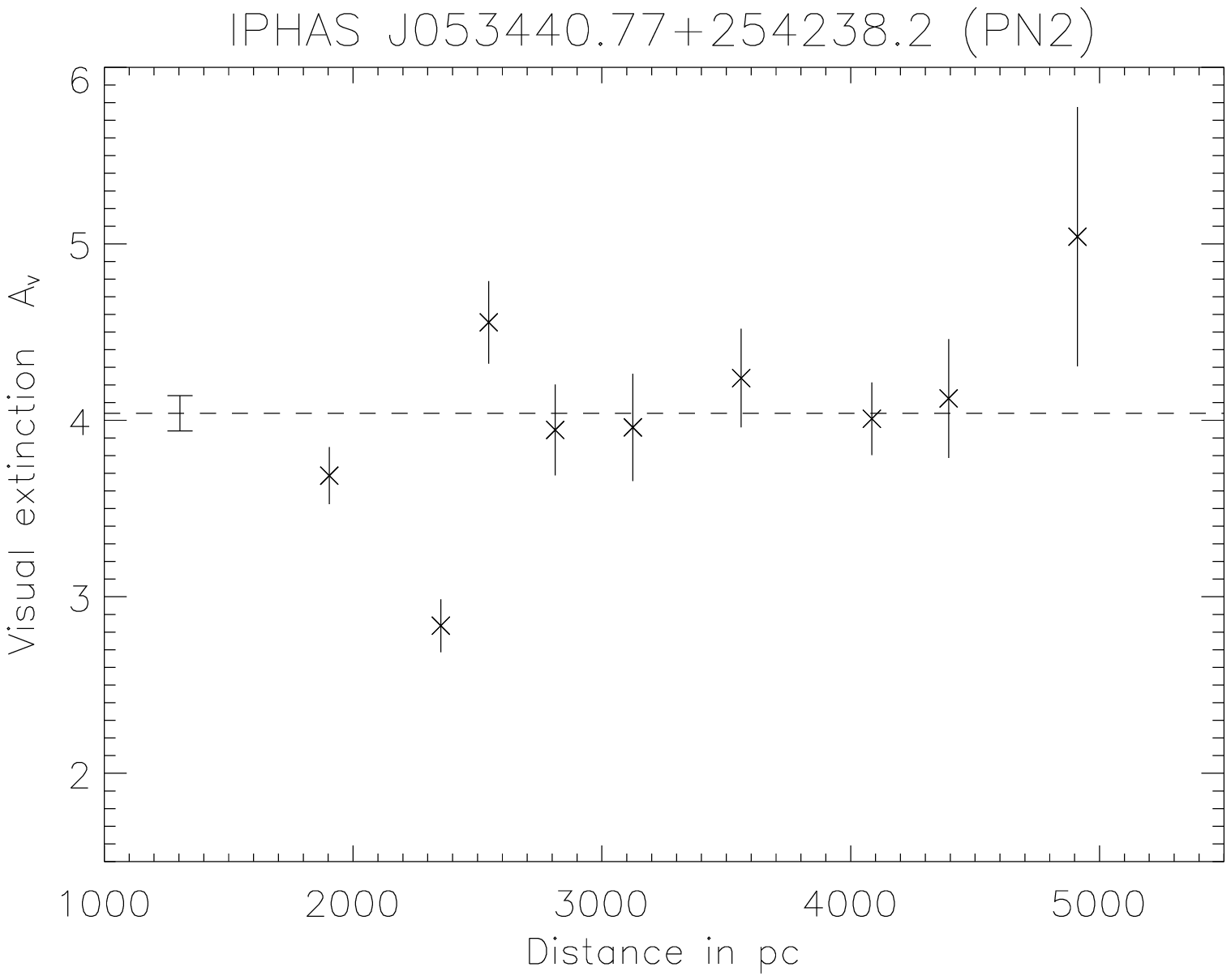}\includegraphics[width=0.5\linewidth,angle=0]{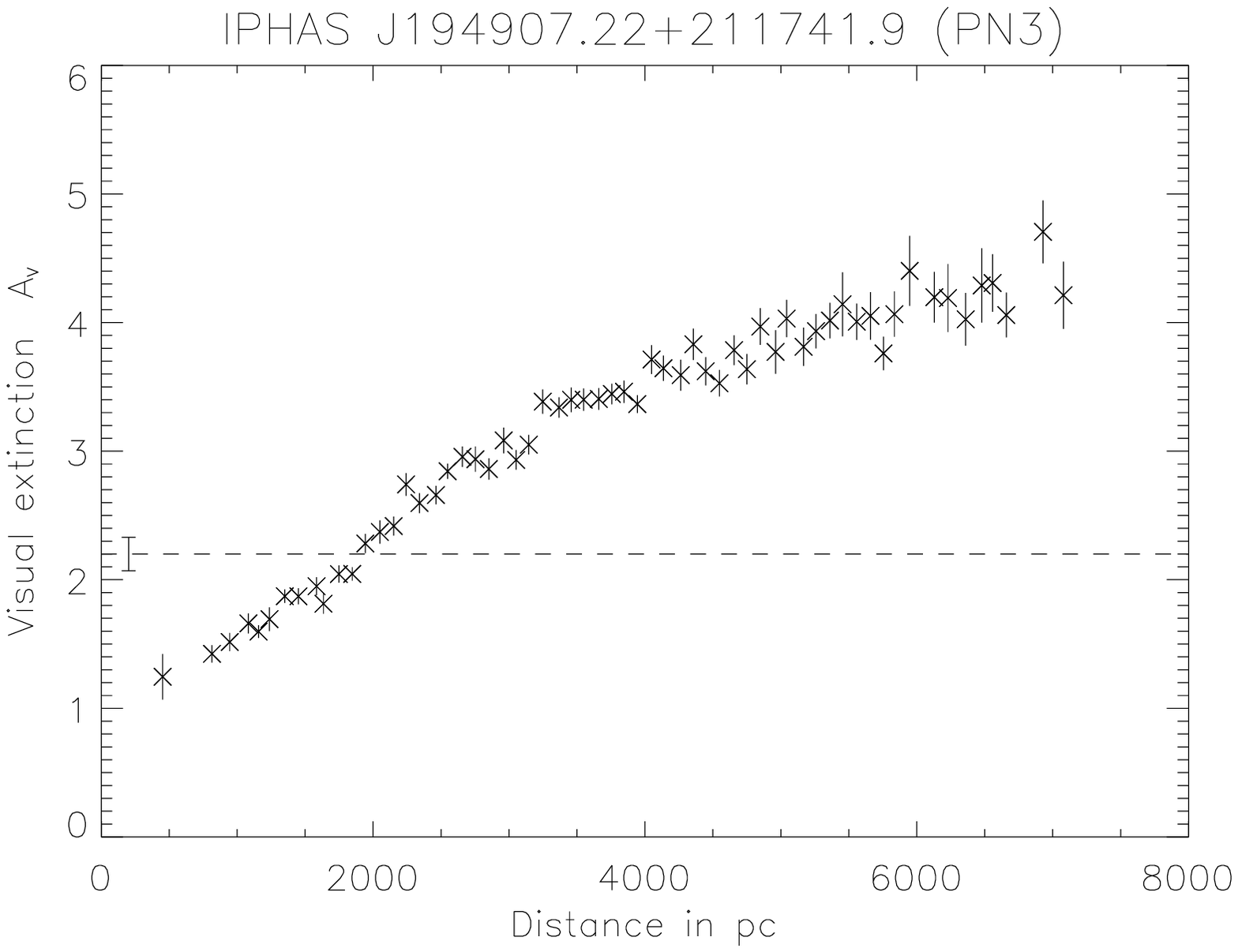}
\includegraphics[width=0.5\linewidth,angle=0]{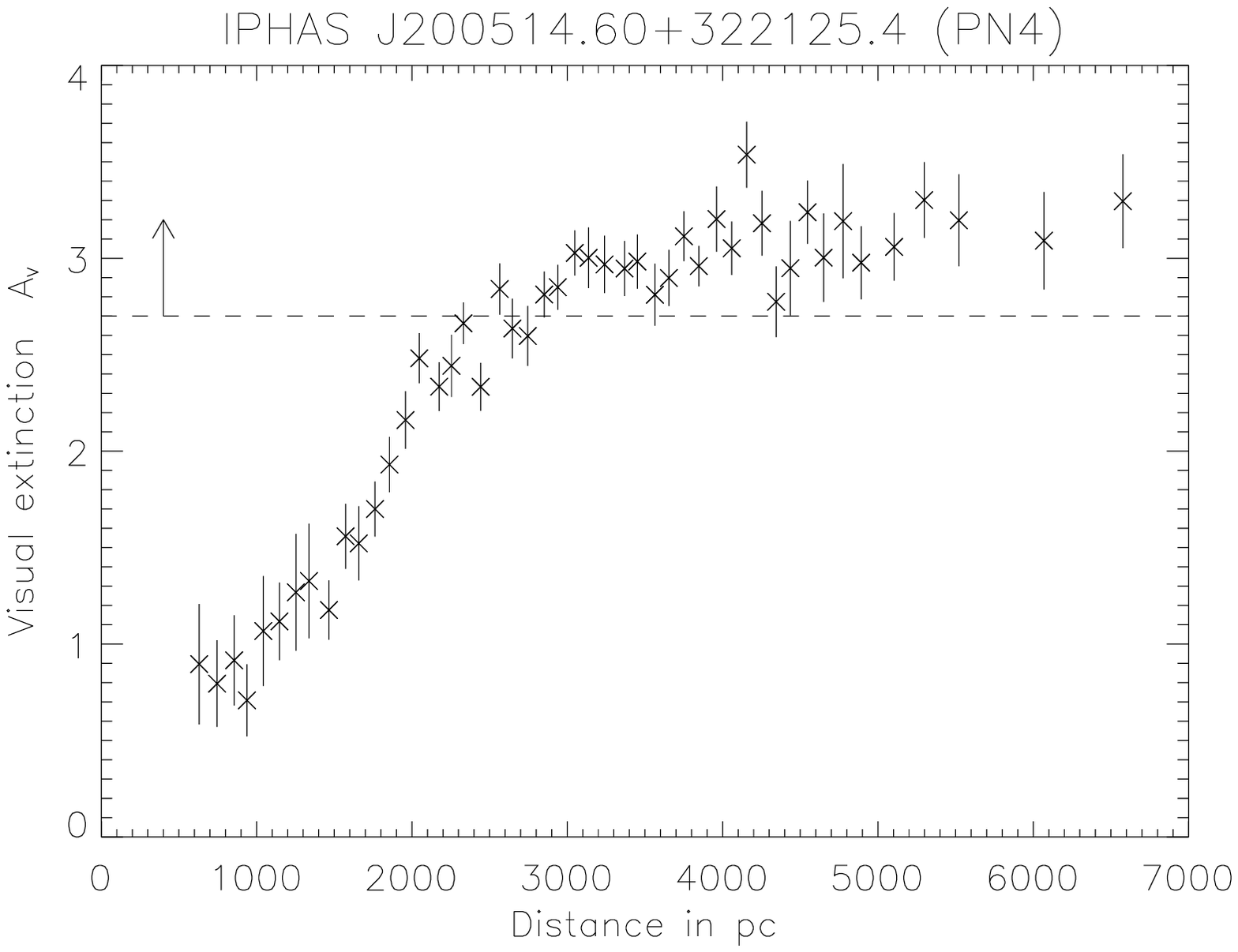}\includegraphics[width=0.5\linewidth,angle=0]{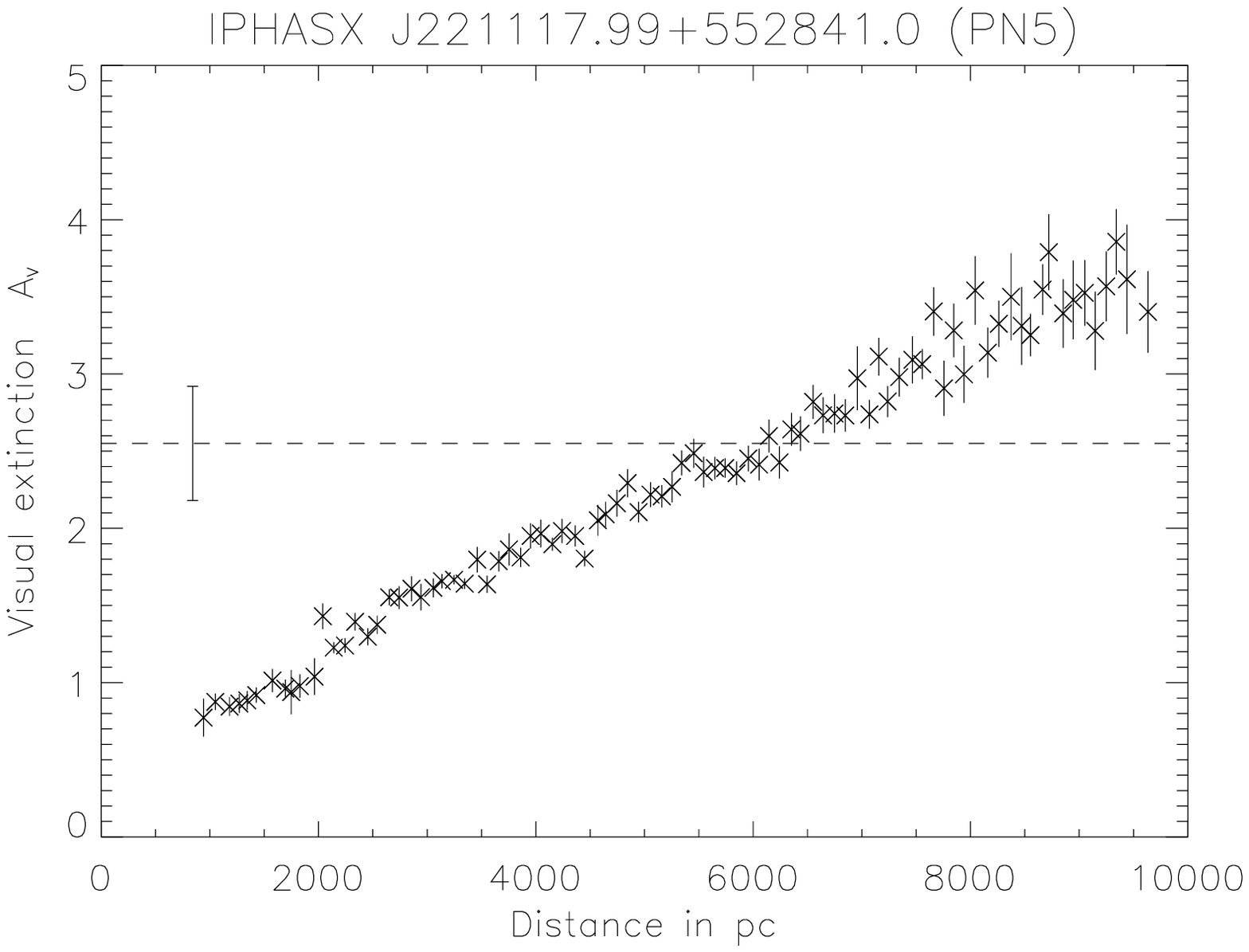}
\protect\caption[ ]{The extinction - distance curves \citep{sale09} for the new IPHAS PNe. Note that for PN4 the estimated distance is only a rough lower limit. Each point refers to binned data given the weighted mean of the extinction and distance for the stars in the bin.  The measured $A_V$ is shown as a dashed line, with the vertical bar showing the measurement error.}
\label{fig:15}
\end{figure*}

\section{Distances}\label{s8}

IPHAS data provide the basis for a new distance determination tool that maps extinction at high angular resolution (a few arcmin), drawing on the photometric parallaxes derived from large numbers of A-K stars (\citealp{sale09}, see also \citealp{drew08}). In Fig.~\ref{fig:15}, the extinction $-$ distance ($A_V - D$) curves for the new PN sightlines, obtained from IPHAS field-star photometry, are plotted together with the visual extinctions and corresponding error bars for each of the new PNe. We note that the possible effect of local extinction is not considered in this study; if it is present for any object, the derived distance will be an overestimate.

For PN1, the curve is not shown since the IPHAS photometry along this sightline is not yet of survey standard in that its photometric calibration is suspect.

PN2 is located towards the galactic anticenter, where the largest galactocentric distances are expected, and this makes PN2 especially interesting. However, the background around PN2 is clearly patchy in the IPHAS images due to the presence of a  small molecular cloud centred 10$\arcmin$ from PN2 (section 5.1). Hence particular care was taken when analyzing its extinction curve. Instead of deriving the curve from field stars located in an area of $10\arcmin \times 10\arcmin$ surrounding the object, as is commonly sufficient, a smaller $5\arcmin \times 5\arcmin$ area was used. This leads to a more sparsely sampled curve, but should in principle provide a systematically more reliable guide to PN2's sightline.

The data for PN2 (Fig.~\ref{fig:15}) show several jumps well above the quoted errors; e.g., at 2.35 and 2.55 kpc. This happens because most stars in those bins are not evenly distributed across the $5 \times 5$ field. For instance, 7 of 9 stars in the 2.55 kpc bin are grouped in the periphery of an apparent ''hole'' (i.e., an area empty of stars) located $\sim50\arcsec$ SE of PN2, and serve to push the extinction in that bin above that of its neighbours. Nevertheless, what we do learn with some confidence is that the reddening of PN2 is comparable with the asymptotic value derived from the field stars, which accordingly places a lower limit on the distance, $D \ga 2$ kpc.

For PN3, the $A_V - D$ curve is well sampled and the distance, $D=1.8~\pm$~0.2 kpc deduced from it, is reliable.

For PN4, only a lower limit to the extinction is available (Sec.~\ref{ext}), and this yields a limit to its distance $D \ga 2.2$ kpc.  However, we note that the implication of the extinction $-$ distance relation for this sightline (Fig.~\ref{fig:15}) is that the interstellar extinction toward PN4 is very likely to be within half a magnitude of $A_V = 3.2$. We used this tentative value in previous sections when needed.

For PN5, the extinction $-$ distance relation implies a distance of 6.1~$\pm$~1.1 kpc. Since this object is very extended, we can also estimate its distance using the empirical H$\alpha$ surface brightness - radius ($S - R$) method by \cite{frew06}. The total H$\alpha$ flux was measured from the IPHAS images and corrected for the contribution of [N\,{\sc ii}] lines using the ratio derived from the spectrum. Ignoring the very faint outer lobes, a reddening corrected total H$\alpha$ flux of $2.2 \pm 0.6 \times 10^{-12}$ erg cm$^{-2}$s$^{-1}$ is obtained. From the $S - R$ relation for a subset of bipolar nebulae given in \cite{frew06}, from an apparent area of the central ellipse of the PN of $1000\pm100$ arcsec$^2$, and from a geometrical radius of the central ellipse of 17$\pm 1\arcsec$, a distance of D = 6.8$\pm$1.2 kpc is derived. As mentioned in \cite{frew06}, estimating the size of bipolar PNe is generally more difficult than for round and elliptical nebulae and this leads to greater surface-brightness errors.
However, it is clear that the distances derived from the $A_V - D$ method and from the $S - R$ method are large and mutually consistent.  Adopting the extinction distance of 6.1 kpc, the linear dimension of the object can be shown to be a quite remarkable 2.2 pc.

\section{Discussion and conclusions}\label{s9}

IPHAS provides us with a massive H$\alpha$ database for the Galactic plane of unprecedented depth and spatial resolution and is therefore able to unveil rare objects (such as those in the symbiotic/PNe uncertain area) or extremely short-lived stages of evolution, such as the transition from a protoPN to PN.

We have presented, in Table~\ref{tab:1}, a list of 83 IPHAS objects from
the W08 catalogue of IPHAS emission-line stars selected as PN
candidates. For 35 candidates, either spectra is obtained or the object type is determined using literature data. Eight objects are
classified as PNe, four of which are newly discovered objects. In
addition, one extended PN (PN5) was found serendipitously; it is a
distant bipolar PN, and its large physical radius implies an old age.

Of the remaining four new objects (PN1, PN2, PN3, and PN4), the last
shows extended bipolar lobes around an unresolved high-density core,
while the other three are unresolved.
These four nebulae have some remarkable characteristics in common: all
are extremely dense, PN1, PN2, and PN4 have iron lines in emission, and
PN1 and PN2 also show the Ca\,{\sc ii} triplet in emission. The presence of those lines generally indicates very high densities, and even evolved PNe hosting an interacting binary nucleus can show  Ca\,{\sc ii} lines in emission, as in the case of the quadrupolar PN G 126.6+1.3 \citep{mampaso06}. In addition to these features, PN1, PN2, PN3, and PN4 also show low-excitation emission-lines typical of PNe.

In Sect.~\ref{s6}, we studied our objects in various diagnostic
diagrams to gain better insight into their nature.
PN3 is the easiest to interpret. Its small apparent size, relatively low
excitation spectrum, high density, and diagnostic diagrams all
support its classification as a young PN. The minimum size of a PN is
set by the time between the end of the AGB and the onset of
ionisation. This is typically $\lesssim10^3$ yr, which for a constant
expansion velocity of 20~kms$^{-1}$ would give a minimum physical radius
of $\lesssim$ 0.02 pc. The physical angular radius of PN3, derived from
the estimated distance, is $\lesssim$0.003 pc. This reinforces its
classification as a very young PN.

The diagnostics of PN1, PN2, and PN4 are not straightforward to
interpret. Their optical colours and position in the diagnostic-line
diagrams that we have investigated, are fully consistent with the
hypothesis that they are very young PNe, still physically very
compact, but already photoionized - at least partially - by the
central star. Their spectra and emission-line ratios do not support 
the alternative hypothesis that they are young stellar objects (see Sect. 7).

However, their colours at wavelengths longer than the optical regime
(2MASS, plus MSX and IRAS when available) are not those typical of the bulk of known PNe. In particular, in the 2MASS colour-colour diagram (Figs. 1 and 2),
they are located in the area occupied by symbiotic stars of the D
(dusty) type. These are interacting binaries containing a hot white
dwarf accreting from the wind of a Mira variable. A large fraction of
them \citep{corradi03} display extended
bipolar nebulae, which are morphologically, spectrally and dynamically
indistinguishable from PNe. Separating the two classes of objects from
an observational point of view is a difficult task \citep{santander08}, and there is an increasing sample of objects
classified as PNe falling in the uncertain category \citep{corradi95}. Some of them are extremely well-studied nebulae, like M2~9 and
Mz~3, which confirms the intrinsic difficulty in ascertaining the
exact nature of these objects. As a further proof of this, we note that,
in spite of its ``symbiotic'' NIR colours, VLTI observations of
Mz~3 indicate that the NIR emission of this object comes from a dusty
disc, too small to be able to accommodate a Mira star
\citep{chesneau07}.  Therefore, some objects classified as PNe and located in the ``uncertain'' zone of the 2MASS diagram can be symbiotic stars, and viceversa. This reminds us that colour-colour diagrams or
diagnostic-line ratios are useful empirical tools for a first
classification of objects, but they just indicate specific physical
conditions that can be shared by objects of differing evolutionary status.

PN4 (and perhaps PN1 and PN2 also) fall in that
uncertain category, and we should keep open the possibility that they
contain a D-type symbiotic binary source.  However, they clearly do not show red-giant molecular bands (TiO, CO, CN, etc.), nor the 6830 and 7082{\AA} OVI Raman scattered emission lines, nor lines from very high excitation ions (up to $>$100eV like Fe$^{+6}$) which are typical of symbiotics. So, they do not fulfil the observational definition of symbiotic stars given by
\citet{belczynski00}, and, as a working hypothesis, we classify them
as very young PNe. Our effort, in the future, will be directed to
determining whether a symbiotic Mira is hidden in their
centre -- cf. the case of He 2-104, whose symbiotic nature is
undetected in the optical and only revealed by NIR spectra and
photometric light modulation \citep{santander08b}.

\begin{acknowledgements}
We thank David Frew for providing his help and needed formula for $S - R$ distance determination and Gerardo Juan Luna for reducing the Chandra data. K.~V., A.~M., and R.~L.~M.~C. acknowledge funding from the Spanish AYA2007-66804 grant. M.~R. and G.~D.-I. acknowledge support from Mexican CONACYT project 50359-F. S.~E.~S. is in receipt of a studentship funded by the Science \& Technology Facilities Council of the United Kingdom. J.~S. acknowledges support from grant SAO G06-7022. This paper makes use of data obtained as part of IPHAS carried out at the INT as well as INT, WHT, NOT and SPM2.1m spectroscopic data. The INT and WHT are operated by the Isaac Newton Group and NOT by NOTSA on the island of La Palma in the Spanish Observatorio del Roque de los Muchachos of the Instituto de Astrof\'{i}sica de Canarias. SPM 2.1m is operated by UNAM at the OAN of SPM, Mexico. All IPHAS data are processed by the Cambridge Astronomical Survey Unit, at the Institute of Astronomy in Cambrid
 ge. We also acknowledge use of data products from the 2MASS, which is a joint project of the University of Massachusetts and the Infrared Processing and Analysis Centre/California Institute of Technology (funded by the USA's National Aeronautics and Space Administration and National Science Foundation).

\end{acknowledgements}

\bibliographystyle{aa}
\bibliography{1157}

\scriptsize
\longtabL{1}{
\begin{landscape}
\begin{longtable}{lllllllllllll}
%\begin{threeparttable}
\caption{\label{tab:1} The PN candidates in W08 selected based on their IPHAS and 2MASS colours (see text).}\\
\hline \hline
Coordinates & $r^\prime$  mag & $r^\prime - i^\prime$ mag & $r^\prime -$ H$\alpha$ mag & $J$ mag & $H$ mag & $K$ mag & FWHM$_{\textrm{stars}}$ & FWHM  & SIMBAD id. & S & Id. & $d_4$\\
 \hline
\endfirsthead
\caption{continued.}\\
\hline\hline
Coordinates & $r^\prime$ mag & $r^\prime - i^\prime$ mag & $r^\prime -$ H$\alpha$ mag & $J$ mag & $H$ mag & $K_s$ mag & FWHM$_{\textrm{stars}}$ & FWHM & SIMBAD id.& S & Id. & $d_4$\\
\hline
\endhead
\hline
00 10 20.53 +58 37 08.6 & 19.363 (0.057) & 1.160 (0.067) & 2.348 (0.058) & 15.739 (0.072) & 14.281 (0.045) & 13.218 (0.034) & 0.97 (0.03) & 0.96& & y & Em* & 7.3 \\ 
00 28 10.82 +61 41 31.2 & 18.850 (0.014) & 1.142 (0.024) & 1.296 (0.021) & 15.997 (0.097) & 15.335 (0.131) & 14.619 (0.085) & 0.95 (0.05) & 0.97& & n & ? & 14.7 \\ 
00 42 54.48 +61 19 31.2 & 16.891 (0.004) & 1.429 (0.006) & 1.562 (0.006) & 12.534 (0.027) & 11.149 (0.033) & 09.984 (0.019) & 0.89 (0.02) & 0.88& IRAS 00399+6103 & n & ? & 4.3 \\ 
00 54 49.44 +63 35 42.6 & 19.259 (0.021) & 1.279 (0.033) & 1.363 (0.028) & 15.340 (0.051) & 14.133 (0.052) & 13.027 (0.030) & 0.83 (0.03) & 0.82& & n & ? & 31.7 \\ 
01 25 44.66 +61 36 11.7 & 18.870 (0.054) & 0.669 (0.069) & 2.009 (0.056) & 15.483 (0.054) & 13.218 (0.030) & 11.492 (0.020) & 1.24 (0.04) & 1.25& IRAS 01224+6120 & y & new PN (PN1) & 21.5 \\ 
01 45 51.20 +64 16 05.5 & 19.026 (0.052) & 1.170 (0.061) & 1.685 (0.055) & 15.648 (0.082) & 14.504 (0.087) & 13.283 (0.052) & 1.43 (0.07) & 1.41& & & ? & 17.5 \\ 
02 00 39.48 +60 32 59.1 & 14.734 (0.003) & 0.615 (0.004) & 1.949 (0.003) & 12.452 (0.023) & 11.263 (0.021) & 09.880 (0.020) & 0.92 (0.02) & 0.95& HBHA 6211-02, IRAS 01571+6018 & n & Em* ? & 33.6 \\ 
02 41 35.93 +57 37 38.0 & 18.817 (0.015) & 0.934 (0.027) & 1.133 (0.022) & 16.423 (0.106) & 15.145 (0.085) & 13.296 (0.039) & 1.01 (0.07) & 1.01& IRAS02379+5724 & n & known YSO & 7.0 \\ 
03 20 39.49 +56 23 58.2 & 14.675 (0.001) & 1.275 (0.002) & 1.607 (0.001) & 10.953 (0.029) & 09.567 (0.033) & 08.216 (0.024) & 1.02 (0.03) & 1.03& IRAS 03168+5613 & y & Em* & 35.3 \\ 
03 21 14.32 +61 05 26.7 & 16.837 (0.005) & 0.539 (0.011) & 1.298 (0.007) & 15.751 (0.078) & 15.553 (0.149) & 14.974 (0.109) & 0.99 (0.03) & 1.01& V* FT Cam. & n & known Nova & 34.2 \\ 
04 10 11.85 +50 59 54.6 & 16.825 (0.004) & 2.120 (0.005) & 1.448 (0.006) & 10.923 (0.022) & 08.902 (0.028) & 07.116 (0.017) & 0.74 (0.03) & 0.78& CPM 12 & n & YSO\footnote{Field with diffuse emission} & 40.8 \\ 
04 22 56.11 +53 47 09.4 & 19.296 (0.024) & 1.096 (0.037) & 1.233 (0.033) & 16.266 (0.096) & 15.566 (0.133) & 14.754 (0.100) & 0.78 (0.04) & 0.72& & n & ? & 24.3 \\ 
04 27 54.38 +48 10 08.6 & 19.345 (0.031) & 1.251 (0.044) & 1.549 (0.039) & 15.506 (0.069) & 13.545 (0.044) & 11.935 (0.030) & 1.65 (0.10) & 1.72& & n & ? & 38.1 \\ 
04 29 26.85 +45 16 10.2 & 17.227 (0.005) & 1.289 (0.009) & 1.488 (0.007) & 14.039 (0.025) & 12.622 (0.027) & 11.435 (0.019) & 0.91 (0.01) & 0.91& & n & ? & 12.2 \\ 
04 43 38.64 +41 09 41.5 & 15.016 (0.002) & 0.995 (0.004) & 1.184 (0.003) & 12.249 (0.025) & 11.194 (0.024) & 10.171 (0.018) & 0.98 (0.03) & 0.99& IRAS 04401+4103 & n & ? & 42.7 \\ 
04 56 25.15 +43 49 31.8 & 18.130 (0.011) & 0.999 (0.019) & 1.843 (0.014) & 15.247 (0.044) & 14.077 (0.032) & 13.035 (0.024) & 1.46 (0.02) & 1.47& & y & Em* & 36.8 \\ 
05 02 19.46 +48 03 37.6 & 18.676 (0.020) & 1.050 (0.039) & 1.326 (0.028) & 16.116 (0.094) & 15.319 (0.106) & 14.440 (0.065) & 1.76 (0.08) & 1.79& & n & ? & 55.4 \\ 
05 22 43.78 +33 25 25.8 & 16.511 (0.004) & 1.418 (0.006) & 1.271 (0.006) & 12.384 (0.023) & 10.771 (0.022) & 09.423 (0.019) & 1.50 (0.02) & 1.48& Cl* NGC 1893 NMI E9 & n & ?\footnotemark[1]{$^{,}$}\footnote{Pre-main sequence Star Candidate} & 4.2 \\ 
05 32 09.27 +30 18 53.0 & 18.676 (0.028) & 1.310 (0.033) & 1.276 (0.032) & 15.199 (0.045) & 14.253 (0.044) & 13.371 (0.034) & 1.31 (0.04) & 1.40& & n & ? & 8.3 \\ 
05 34 40.77 +25 42 38.2 & 17.477 (0.010) & 1.197 (0.013) & 1.844 (0.011) & 14.088 (0.027) & 12.122 (0.027) & 10.584 (0.021) & 0.73 (0.03) & 0.73& & y & new PN (PN2) & 11.3 \\ 
05 38 17.88 +31 39 34.8 & 16.230 (0.005) & 1.036 (0.006) & 1.347 (0.006) & 13.464 (0.023) & 12.630 (0.023) & 11.713 (0.020) & 0.76 (0.03) & 0.75& & n & ? & 22.5 \\ 
06 00 04.54 +16 51 26.0 & 18.975 (0.016) & 0.800 (0.031) & 1.553 (0.021) & 16.690 (0.152) & 15.687 (0.147) & 14.707 (0.104) & 1.33 (0.05) & 1.35& & n & ? & 28.8 \\ 
06 05 15.15 +20 40 36.7 & 17.820 (0.012) & 1.339 (0.019) & 1.391 (0.016) & 14.772 (0.034) & 13.208 (0.023) & 11.217 (0.020) & 1.91 (0.20) & 1.75& IRAS 06022+2040 & y & Em* & 56.1 \\ 
06 09 26.79 +24 55 19.9 & 16.483 (0.008) & 0.551 (0.016) & 1.061 (0.013) & 12.263 (0.023) & 11.204 (0.022) & 10.244 (0.017) & 1.29 (0.02) & 1.29& & n & ? & 39.3 \\ 
06 13 17.53 +15 19 57.8 & 19.479 (0.025) & 1.136 (0.045) & 1.355 (0.033) & 16.073 (0.086) & 15.291 (0.084) & 14.293 (0.070) & 1.25 (0.03) & 1.28& & n & ? & 4.6 \\ 
06 14 17.27 +22 54 18.6 & 18.596 (0.013) & 0.490 (0.032) & 1.900 (0.016) & 14.552 (0.035) & 13.294 (0.033) & 12.277 (0.024) & 1.09 (0.02) & 1.09& & n & ? & 22.4 \\ 
06 20 21.71 +11 07 10.2 & 18.622 (0.019) & 1.175 (0.027) & 1.185 (0.026) & 16.220 (0.121) & 15.075 (0.093) & 13.992 (0.073) & 1.08 (0.05) & 1.12& IRAS 06175+1108 & n & ? & 40.1 \\ 
06 24 23.36 +09 24 17.5 & 18.512 (0.011) & 1.002 (0.019) & 1.182 (0.017) & 15.626 (0.065) & 14.962 (0.063) & 14.159 (0.065) & 0.89 (0.02) & 0.86& & n & ? & 13.2 \\ 
06 39 34.25 +06 21 16.9 & 18.805 (0.023) & 0.498 (0.044) & 1.323 (0.031) & 16.768 (0.173) & 15.786 (0.129) & 14.786 (0.113) & 1.33 (0.18) & 0.97& & n & ? & 13.0 \\ 
06 42 22.19 -02 26 28.6 & 19.224 (0.028) & 1.449 (0.041) & 1.280 (0.038) & 15.851 (0.070) & 14.899 (0.073) & 13.754 (0.050) & 1.42 (0.04) & 1.39& & n & ? & 26.2 \\ 
18 35 01.83 +01 46 56.0 & 16.344 (0.004) & 1.875 (0.005) & 2.516 (0.004) & 10.745 (0.021) & 09.476 (0.025) & 08.921 (0.025) & 1.33 (0.03) & 1.32& & y & symbiotic star\footnote{Also in \cite{corradi08} symbiotic star candidate list} & 26.4 \\ 
18 38 07.38 +00 11 13.6 & 18.466 (0.011) & 1.386 (0.016) & 2.351 (0.013) & 14.195 (0.037) & 13.239 (0.032) & 12.793 (0.032) & 1.21 (0.25) & 1.01& & y & Em*\footnotemark[3] & 36.0 \\ 
18 50 05.71 -00 40 41.2 & 16.951 (0.004) & 1.364 (0.006) & 1.228 (0.006) & 13.001 (0.030) & 11.873 (0.027) & 10.655 (0.021) & 1.17 (0.04) & 1.08& & n & ? & 36.1 \\ 
18 51 41.55 +09 54 52.4 & 15.342 (0.002) & -0.124 (0.005) & 2.799 (0.002) & 14.457 (0.040) & 13.909 (0.062) & 12.712 (0.025) & 0.93 (0.02) & 0.97& PN K 3-15 & n & known PN & 59.5 \\ 
18 57 04.44 +00 26 31.7 & 18.280 (0.010) & 2.253 (0.012) & 2.690 (0.011) & 10.425 (0.021) & 08.863 (0.025) & 08.086 (0.021) & 1.00 (0.04) & 1.01& & y & symbiotic star\footnotemark[3] & 24.4 \\ 
19 00 34.82 -02 11 58.0 & 15.915\ (0.003) & 0.521\ (0.006) & 2.915\ (0.003) & 13.293\ (U) & 12.298\ (0.067) & 10.505\ (U) & 1.18 (0.02) & 1.35& PN K 3-18 & n & known PN\footnote{Resolved in the IPHAS H$\alpha$ image}{$^{,}$}\footnote{U = 2MASS assignation for upper limit on magnitude} & 27.7 \\ 
19 02 29.97 -02 27 57.0 & 16.128 (0.003) & -0.168 (0.009) & 2.937 (0.004) & 13.416 (0.036) & 11.222 (0.029) & 09.343 (0.024) & 1.70 (0.05) & 1.71& & y & Em* & 12.8 \\ 
19 10 17.43 +06 52 58.1 & 15.815 (0.003) & 1.926 (0.004) & 1.719 (0.004) & 10.416 (0.023) & 08.708 (0.022) & 07.224 (0.020) & 0.96 (0.02) & 0.86& IRAS 19078+0647 & n & ? & 84.3 \\ 
19 12 33.23 +11 46 31.2 & 18.598 (0.036) & 1.739 (0.043) & 1.505 (0.043) & 13.995 (0.032) & 12.668 (0.030) & 10.919 (0.023) & 0.99 (0.08) & 0.98& & n & ? & 33.2 \\ 
19 17 01.33 +15 59 47.8 & 17.697 (0.007) & 1.492 (0.011) & 1.354 (0.011) & 13.426 (0.046) & 12.056 (0.044) & 10.762 (0.033) & 0.90 (0.03) & 0.85& & n & ? & 27.0 \\ 
19 17 50.56 +08 15 08.5 & 15.210 (0.002) & 0.619 (0.004) & 2.626 (0.002) & 12.981 (0.028) & 12.556 (0.037) & 11.925 (0.026) & 0.91 (0.02) & 0.95& J191750.56+081508.5 & n & known possible PN & 93.2 \\ 
19 22 26.67 +10 41 21.2 & 17.219 (0.006) & 0.223 (0.015) & 2.982 (0.007) & 14.340 (0.022) & 13.794 (0.046) & 12.805 (0.035) & 1.84 (0.08) & 1.88& PN K 3-33 & n & known PN & 33.2 \\ 
19 22 49.80 +14 22 36.3 & 16.957 (0.005) & 1.635 (0.007) & 1.339 (0.007) & 12.614 (0.024) & 11.189 (0.028) & 09.877 (0.018) & 1.78 (0.16) & 1.79& & n & ? & 66.5 \\ 
19 27 17.94 +08 14 29.4 & 17.201 (0.006) & 0.171 (0.017) & 0.918 (0.010) & 15.317 (0.076) & 13.995 (0.055) & 12.763 (0.036) & 1.43 (0.04) & 1.41& & n & ? & 44.4 \\ 
19 28 02.95 +17 16 43.3 & 19.251 (0.024) & 1.680 (0.033) & 1.717 (0.028) & 14.168 (0.035) & 12.575 (0.039) & 10.882 (0.030) & 0.92 (0.03) & 0.99& IRAS 19258+1710 & n & ? & 35.5 \\ 
19 44 05.25 +23 26 47.9 & 14.046 (0.001) & 0.511 (0.002) & 2.262 (0.001) & 11.772 (0.024) & 10.304 (0.028) & 09.081 (0.020) & 1.06 (0.03) & 1.06& & n & ? & 8.7 \\ 
19 46 07.52 +22 31 12.3 & 17.538 (0.006) & 1.494 (0.009) & 2.835 (0.007) & 09.223 (0.022) & 07.194 (0.036) & 05.966 (0.021) & 1.24 (0.02) & 1.25& & y & symbiotic star & 33.2 \\ 
19 49 07.23 +21 17 42.0 & 16.770 (0.005) & 0.778 (0.009) & 1.718 (0.006) & 13.583 (0.025) & 11.884 (0.029) & 10.494 (0.022) & 1.14 (0.05) & 1.16& & y & new PN (PN3) & 39.2 \\ 
19 54 00.62 +33 22 12.5 & 15.513 (0.005) & -0.050 (0.014) & 2.646 (0.005) & 14.355 (0.031) & 14.028 (0.047) & 13.477 (0.040) & 1.01 (0.07) & 1.04& PN K 3-49 & n & known PN & 34.2 \\ 
19 59 35.55 +28 38 30.3 & 15.810 (0.003) & 0.861 (0.005) & 1.230 (0.004) & 12.478 (0.023) & 10.322 (0.021) & 08.267 (0.021) & 1.10 (0.02) & 1.09& & y & Em* & 20.1 \\ 
19 59 56.42 +30 48 23.8 & 17.478 (0.009) & 1.036 (0.013) & 1.577 (0.011) & 13.688 (0.023) & 12.164 (0.023) & 10.908 (0.019) & 1.51 (0.05) & 1.62& & n & ? \footnotemark[1] & 7.0 \\ 
20 05 14.59 +32 21 25.1 & 17.241 (0.006) & -0.391 (0.019) & 3.058 (0.007) & 14.956 (0.061) & 12.759 (0.032) & 10.656 (0.026) & 1.38 (0.05) & 1.35& IRAS 20032+3212 & y & new PN (PN4)\footnotemark[1] & 15.3 \\ 
20 08 11.53 +35 25 00.5 & 15.577 (0.003) & 0.577 (0.006) & 1.395 (0.004) & 11.315 (0.020) & 10.605 (0.016) & 09.775 (0.019) & 1.47 (0.05) & 1.52& & n & ? & 8.8 \\ 
20 09 19.02 +39 48 52.8 & 17.663\ (0.009) & 0.040\ (0.024) & 2.692\ (0.010) & -- & -- & -- & 1.10\ (0.02) & 1.09 (0.03) & 1.092005 & n & known Nova & 23.5 \\ 
20 11 26.20 +33 16 06.8 & 18.926 (0.018) & 1.374 (0.028) & 1.343 (0.026) & 14.335 (0.029) & 12.936 (0.022) & 11.704 (0.020) & 1.09 (0.06) & 1.02& & n & ? & 19.1 \\ 
20 15 50.96 +37 30 04.2 & 18.070 (0.014) & 1.672 (0.018) & 2.156 (0.016) & 13.079 (0.022) & 11.464 (0.016) & 09.955 (0.016) & 1.85 (0.09) & 1.76& & n & ? & 8.6 \\ 
20 16 23.18 +37 37 04.6 & 19.345 (0.037) & 2.015 (0.043) & 1.536 (0.044) & 14.064 (0.035) & 12.659 (0.027) & 11.480 (0.019) & 1.94 (0.11) & 1.91& & n & ? & 9.9 \\ 
20 18 35.83 +40 55 08.0 & 17.242 (0.007) & 1.025 (0.012) & 1.629 (0.009) & 14.748 (0.036) & 13.814 (0.037) & 12.885 (0.031) & 1.00 (0.03) & 1.11& & n & ? \footnotemark[1]{$^{,}$}\footnotemark[4] & 3.7 \\ 
20 20 58.52 +38 09 49.8 & 16.199 (0.006) & 1.526 (0.007) & 1.600 (0.007) & 11.880 (0.022) & 10.384 (0.018) & 08.865 (0.014) & 1.04 (0.03) & 1.05& & n & ? & 14.2 \\ 
20 25 49.86 +41 23 08.1 & 18.893 (0.014) & 0.834 (0.029) & 1.905 (0.018) & 15.416 (0.052) & 13.848 (0.036) & 12.299 (0.023) & 0.78 (0.06) & 0.86& & n & ?\footnote{Star near a dark lane} & 17.9 \\ 
20 28 34.25 +35 54 17.4 & 15.890 (0.003) & 1.473 (0.004) & 2.081 (0.004) & 11.498 (0.022) & 10.076 (0.021) & 08.295 (0.015) & 1.06 (0.02) & 1.08& CPM P12 & y & known YSO & 45.6 \\ 
20 29 47.93 +35 59 26.5 & 14.754 (0.001) & 1.010 (0.002) & 1.774 (0.001) & 10.963 (0.022) & 09.440 (0.017) & 08.139 (0.015) & 1.06 (0.02) & 2.01& HBHA 3704-06, IRAS 20278+3549 & n & known YSO \footnotemark[4]{$^{,}$}\footnote{Also [KLT2003] SMM 3} & 48.1 \\ 
20 34 13.39 +41 01 57.9 & 18.999 (0.016) & 1.246 (0.026) & 2.462 (0.018) & 13.229 (0.022) & 11.289 (0.018) & 09.860 (0.015) & 0.82 (0.04) & 0.81& & y & Em* & 20.5 \\ 
20 46 45.49 +41 06 59.6 & 18.539 (0.028) & 1.077 (0.033) & 1.396 (0.031) & 15.614 (0.054) & 14.222 (0.040) & 13.030 (0.032) & 0.95 (0.04) & 0.95& & n & ? & 21.8 \\ 
20 47 13.69 +46 35 17.5 & 14.460 (0.002) & 1.608 (0.002) & 1.325 (0.002) & 10.049 (0.027) & 08.925 (0.033) & 07.480 (0.024) & 1.03 (0.05) & 1.02& & n & ? & 14.6 \\ 
20 54 07.40 +41 34 58.0 & 15.119 (0.003) & 1.353 (0.004) & 1.346 (0.004) & 14.622 (0.033) & 12.822 (0.021) & 11.279 (0.017) & 1.64 (0.26) & 1.49& V* V1219 Cyg & n & Variable Star\footnote{Faint H$\alpha$ nebula around the star and small new nebula at $\sim$ 70\arcsec~NW.} & 40.6 \\ 
20 58 02.67 +46 35 02.6 & 19.047 (0.042) & 1.314 (0.048) & 1.251 (0.047) & 15.623 (0.061) & 14.337 (0.037) & 13.201 (0.044) & 0.75 (0.03) & 0.83& & n & ? & 14.2 \\ 
21 04 04.87 +53 51 24.4 & 18.208 (0.013) & 1.577 (0.018) & 2.393 (0.014) & 13.432 (0.025) & 12.401 (0.025) & 11.613 (0.025) & 0.69 (0.02) & 0.69& Star [MK97] 33 & y & Em* & 54.6 \\ 
21 19 38.96 +46 49 13.6 & 18.841 (0.016) & 1.211 (0.026) & 1.694 (0.021) & 15.059 (0.041) & 13.556 (0.030) & 12.293 (0.023) & 1.25 (0.04) & 1.27& & n & ? \footnote{Small group of H$\alpha$ nebulae at $\sim$ 40\arcsec~NE.} & 1.5 \\ 
21 19 56.24 +51 47 04.2 & 18.971 (0.017) & 1.460 (0.026) & 2.054 (0.020) & 15.195 (0.050) & 14.029 (0.035) & 13.010 (0.030) & 1.16 (0.04) & 1.09& & y & Em* & 5.0 \\ 
21 29 55.57 +55 39 04.1 & 17.808 (0.008) & 1.283 (0.012) & 1.654 (0.011) & 14.191 (0.031) & 12.745 (0.035) & 11.382 (0.021) & 0.86 (0.04) & 0.87& IRAS 21283+5525 & n & ? & 26.7 \\ 
21 39 58.25 +50 14 21.0 & 13.242 (0.001) & 0.788 (0.001) & 1.151 (0.001) & 10.942 (0.030) & 09.209 (0.040) & 06.848 (0.018) & 1.12 (0.03) & 1.18& V* V645 Cyg & n & YSO\footnote{Known reflection nebula (Duck neb.)} & 28.2 \\ 
21 46 25.99 +57 28 28.9 & 18.038 (0.012) & 1.080 (0.018) & 2.361 (0.014) & 14.069 (0.034) & 13.064 (0.033) & 12.301 (0.026) & 0.82 (0.03) & 0.83& & n & ? & 12.0 \\ 
21 53 31.40 +56 29 27.0 & 18.792 (0.022) & 0.661 (0.044) & 2.325 (0.024) & 14.944 (0.059) & 13.302 (0.041) & 12.058 (0.026) & 1.22 (0.04) & 1.20& & n & ? & 19.0 \\ 
21 56 28.47 +57 14 45.5 & 15.142 (0.002) & 0.765 (0.004) & 1.659 (0.003) & 12.967 (0.024) & 11.834 (0.031) & 10.314 (0.024) & 0.95 (0.03) & 0.95& HBHA 5704-02 & n & Em*? & 24.5 \\ 
22 01 08.17 +55 54 41.4 & 16.144 (0.004) & 0.780 (0.007) & 1.155 (0.006) & 13.027 (0.022) & 11.615 (0.031) & 10.431 (0.018) & 1.43 (0.07) & 1.49& & n & ? & 11.0 \\ 
22 06 48.59 +54 11 47.5 & 18.672 (0.016) & 1.252 (0.029) & 1.454 (0.021) & 16.484 (0.153) & 15.440 (0.143) & 14.458 (0.112) & 0.95 (0.04) & 0.95& & n & ? & 20.3 \\ 
22 15 11.99 +57 52 51.4 & 17.833 (0.010) & 0.750 (0.019) & 1.690 (0.012) & 14.983 (0.033) & 13.468 (0.031) & 11.969 (0.022) & 1.04 (0.01) & 1.06& & n & ? & 14.3 \\ 
22 15 18.87 +58 23 12.6 & 18.998 (0.019) & 1.360 (0.028) & 1.696 (0.024) & 14.856 (0.048) & 13.222 (0.042) & 11.624 (0.023) & 1.03 (0.03) & 1.04& & n & ? & 16.5 \\ 
22 18 08.49 +56 05 53.3 & 13.726 (0.001) & 0.703 (0.002) & 1.188 (0.002) & 11.377 (0.027) & 10.182 (0.033) & 09.031 (0.019) & 0.95 (0.01) & 1.19& HBHA 5704-09, IRAS 22162+5550 & n & ?\footnotemark[1]{$^{,}$}\footnotemark[4] & 8.1 \\ 
22 26 28.63 +61 20 49.0 & 18.676 (0.059) & 0.543 (0.118) & 2.131 (0.060) & 14.618 (0.040) & 13.748 (0.044) & 13.422 (0.043) & 1.05 (0.03) & 1.06& & y & Em*\footnotemark[3] & 24.2 \\ 
23 17 35.92 +63 45 06.4 & 16.084 (0.003) & 1.469 (0.004) & 1.787 (0.004) & 12.197 (0.024) & 10.719 (0.027) & 09.395 (0.020) & 1.16 (0.03) & 1.13& HBHA 6207-18, IRAS 23154+6328 & n & YSO? & 17.3 \\ 
23 41 25.21 +65 40 42.6 & 15.034 (0.002) & 1.645 (0.003) & 1.482 (0.003) & 10.498 (0.023) & 08.818 (0.028) & 07.454 (0.020) & 1.20 (0.03) & 1.21& HBHA 6707-04, CPM 40 & n & known YSO\footnote{Cometary nebula displaced from the star} & 50.9 \\
\end{longtable}
\end{landscape}
%\end{threeparttable}
}

%\appendix

\end{document}